\documentclass[usegraphicx,usenatbib,twocolumn]{mn2e}

\usepackage{graphicx}
\usepackage{eucal}
\usepackage[dvips]{epsfig}
\usepackage{amssymb}
\usepackage{amsmath}
\usepackage{bm}

\newcommand{\be}{\begin{eqnarray}}
\newcommand{\ee}{\end{eqnarray}}
\newcommand{\bse}{\begin{subequations}}
\newcommand{\ese}{\end{subequations}}

\newcommand{\mrm}[1]{\mathrm{#1}}
\newcommand{\mbb}[1]{\mathbb{#1}}

\newcommand{\mcal}[1]{\mathcal{#1}}

\newcommand{\vect}[1]{\boldsymbol{#1}}
\newcommand{\matr}[1]{\boldsymbol{\mathsf{#1}}}
\newcommand{\tens}[1]{\boldsymbol{\mathsf{#1}}}

\newcommand{\parder}[2]{\frac{\partial #1}{\partial #2}}
\newcommand{\totder}[2]{\frac{\mrm{d} #1}{\mrm{d} #2}}

\newcommand{\diff}[2][]{\mrm{d}^{#1}{#2}\,}

\newcommand{\tr}{\mrm{tr}\,}

\newcommand{\pdf}{\mrm{pdf}}
\newcommand{\cdf}{\mrm{cdf}}

\newcommand{\arcsecs}{\ensuremath{\mathrm{arcsec}}}
\newcommand{\kpc}{\ensuremath{\mathrm{kpc}}}
\newcommand{\Mpc}{\ensuremath{\mathrm{Mpc}}}
\newcommand{\Gpc}{\ensuremath{\mathrm{Gpc}}}
\newcommand{\Myr}{\ensuremath{\mathrm{Myr}}}
\newcommand{\Msolar}{\ensuremath{\mathrm{M}_\odot}}

\newcommand{\zS}{z^\mrm{S}}
\newcommand{\zL}{z^\mrm{L}}
\newcommand{\rE}{r_\mrm{E}}
\newcommand{\IIorIII}{\ensuremath{\mrm{II}\vee\mrm{III}}}

\newcommand\aj{{AJ}}  
\newcommand\apj{{ApJ}} 
\newcommand\apjl{{ApJ}} 
\newcommand\apjs{{ApJS}} 
\newcommand\aap{{A\&A}} 
\newcommand\aaps{{A\&AS}} 
\newcommand\mnras{{MNRAS}} 
\newcommand\prd{{Phys.~Rev.~D}} 
\newcommand\nat{{Nature}} 

\begin{document}

\title[Strong lensing optical depths in a $\Lambda$CDM universe]{Strong lensing optical depths in a $\Lambda$CDM universe}

\author[S. Hilbert, S.D.M. White, J. Hartlap \& P. Schneider]{
Stefan Hilbert,$^1$\thanks{\texttt{hilbert@mpa-garching.mpg.de}}
Simon D.M. White,$^1$
Jan Hartlap,$^2$
and Peter Schneider$^2$
\\$^1$Max-Planck-Institut f{\"u}r Astrophysik,Karl-Schwarzschild-Stra{\ss}e 1, D-85741, Garching, Germany
\\$^2$Argelander-Institut f{\"u}r Astronomie, Auf dem H{\"u}gel 71, D-53121 Bonn, Germany
}

\date{\today}

\maketitle

\begin{abstract}
We investigate strong gravitational lensing in the concordance $\Lambda$CDM
cosmology by carrying out ray-tracing along past light cones through the
Millennium Simulation, the largest simulation of cosmic structure formation
ever carried out. We extend previous ray-tracing methods in order to take full
advantage of the large volume and the excellent spatial and mass resolution of
the simulation. As a function of source redshift we evaluate the probability
that an image will be highly magnified, will be highly elongated or will be
one of a set of multiple images. We show that such strong lensing events can
almost always be traced to a single dominant lensing object and we study the
mass and redshift distribution of these primary lenses. We fit analytic models
to the simulated dark halos in order to study how our optical depth
measurements are affected by the limited resolution of the simulation and of
the lensing planes that we construct from it. We conclude that such effects
lead us to underestimate total strong-lensing cross sections by about 15\%.
This is smaller than the effects expected from our neglect of the baryonic
components of galaxies.  Finally we investigate whether strong lensing is
enhanced by material in front of or behind the primary lens.  Although strong
lensing lines-of-sight are indeed biased towards higher than average mean
densities, this additional matter typically contributes only a few percent of
the total surface density.
\end{abstract}


\section{Introduction}
\label{sec:Introduction}

Gravitational lensing was first discovered through strong lensing effects
which can produce multiple images of distant quasars~\citep{WalshEtal79} and highly
distorted images of distant extended objects such as galaxies~\citep{LyPe86,SoucailEtal87} and radio sources~\citep{HewittEtal88}. Such effects occur when the surface mass density of an individual object (the `lens') is comparable to
that across the Universe as a whole, and as a result they are generated only
by the most massive and most concentrated structures. In contrast, weak
lensing, detected through the small but coherent distortion of the images of
distant galaxies in the same direction on the sky~\citep{TysonEtal90}, is sensitive to the abundance and structure of typical nonlinear
objects, so-called dark matter halos, and is now beginning to measure the
statistics of the cosmic mass distribution also in the quasilinear regime~\citep{SemboloniEtal06,HoekstraEtal06,HetterscheidtEtal07,MasseyEtal07astroph}. Thus gravitational lensing complements microwave
background, large-scale structure, and Ly$\alpha$ forest studies, which
provide information primarily in the quasilinear and linear regimes \citep[for recent reviews on strong and weak lensing, see][]{Kochanek06_SaasFee33Part2,Schneider06_SaasFee33Part3}. Combining
all these measures to constrain theories for the origin of structure
requires a reliable model for the nonlinear phases of evolution. 

The current standard model of cosmological structure formation is based on
cold dark matter and a cosmological constant. This $\Lambda$CDM
model has been shown to fit a wide variety of observations of galaxies and
their dark halos, of galaxy clusters and galaxy clustering, of the structure
of the intergalactic medium out to redshift 6, and, most notably, of the
detailed pattern of temperature fluctuations in the cosmic microwave
background. The parameters of the model are already highly
constrained~\citep{SpergelEtal07_WMAP_3rdYearData}. Early predictions for its
lensing properties were based on analytic models for nonlinear structure
 \citep[e.g.][]{TurnerEtal84,SubramanianEtal87,NaWh88}, but reliable predictions
require numerical simulation \citep[e.g.][]{BartelmannEtal98,JaSeWh00,WaBoOs04}.
Recent gravitational lensing work has confirmed the dark halo structure
predicted by these simulations for galaxies \citep{HoekstraEtal02,SeljakEtal05,MandelbaumEtal06_Galaxies,SimonEtal07} and
clusters \citep{ComerfordEtal06,MandelbaumEtal06_GroupsAndClusters,NatarajanEtal07} as well as for the ensemble
properties of the dark halo population \citep{SemboloniEtal06,HoekstraEtal06}. Substantial efforts
are currently underway to improve all these measurements, and these will need
to be matched by a comparable improvement in the theoretical predictions.

Additional tests and constraints may be obtained from observations of
gravitational lensing effects.  For example, foreground matter inhomogeneities
may (de-)magnify images of distant sources thus changing their apparent
magnitude. Although expected to be small for current type Ia supernova samples
\citep{WaCeXuOs97,Holz98,RiessEtal98,Valageas00,AmMoGo03,KnopEtal03,BarrisEtal04,RiessEtal04},
some evidence for lensing effects on the observed luminosity distribution may
be present in higher-redshift samples \citep{Wang05,JoenssonEtal06}. For
future supernova surveys, one should be able to detect such effects with
higher statistical significance \citep{Metcalf99,MeSi99,MiHeHa02,MuVa06astroph}. This will then provide a
further test of the standard structure formation model.

Sufficiently massive and concentrated structures along the line-of-sight can
give rise to multiple images, to strongly magnified images and to highly
distorted images, so-called giant arcs. The number of such strong-lensing
events depends on the abundance of massive objects and on their detailed
internal structure, both of which are sensitive to the background
cosmology. Currently a much-debated question is whether the observed frequency
of giant arcs \citep{LuppinoEtal99,ZaGo03,GladdersEtal03} is too high compared to
predictions based on $\Lambda$CDM models with parameters favoured by other
observations \citep{BartelmannEtal98,MeneghettiEtal00,OgLeSu03,DaHoHe04,WaBoOs04,LiEtal05,HoreshEtal05,LiEtal06,WuCh06,MeneghettiEtal07}.
The problem appears particularly pressing at higher redshift, but available
simulation results are not good enough to establish a clear discrepancy.

In this paper, we will study the optical depth for a variety of strong lensing
effects as a function of source redshift in the standard $\Lambda$CDM
cosmology. In particular, we will estimate the fraction of images that are
highly magnified by gravitational lensing, that have a large length-to-width
ratio, or that belong to multiply imaged sources. In addition, we compare the
effect of foreground and background matter to that of the primary lens in
generating these optical depths.

The results presented here were obtained by shooting random rays through a
series of lens planes created from the Millennium
Simulation~\citep{SpringelEtAl05MSReview}. This very large $N$-body simulation
of cosmic structure formation covers a volume comparable to the largest
current surveys with substantially better resolution than previous simulations
used for ray-tracing studies. Our set of lens planes represents the entire
mass distribution between source and observer, allowing us to quantify the
effects of foreground and background matter. In addition we have ensured that
our ray-tracing techniques take full advantage both of the statistical power
offered by the large volume of the simulation, and of its high spatial and
mass resolution. This allows us to make more precise statements about model
expectations than has previously been possible.

Our paper is organised as follows.  In Sec.~\ref{sec:methods}, we describe how
we trace representative rays through the Millennium Simulation. Results for
the magnification distribution as a function of source redshift are presented
in Sec.~\ref{sec:magnification_distribution}. Strong-lensing optical depths
are then discussed in Sec.~\ref{sec:optical_depths}. In
Sec.~\ref{sec:lens_properties}, the mass and redshift distribution of the
objects which cause strong lensing are examined, and we demonstrate that the
errors induced by the finite volume and resolution of the simulation are
relatively small. Biases induced by additional structure in front of or behind
the principal lens are examined in Sec.~\ref{sec:los_matter} and are also
found to be small. The paper concludes with a summary and outlook in
Sec.~\ref{sec:summary}.

\section{Methods}
\label{sec:methods}
\subsection{The multiple-lens-plane approximation}
\label{sec:MLA}
In the multiple-lens-plane approximation \citep[see, e.g.,][]{BlNa86,ScEhFa92}, a
finite number of planes are introduced along the line of sight, onto which the
matter inhomogeneities in the backward light cone of the observer are
projected. Between these lens planes, light is assumed to travel on straight
lines. Light rays are deflected only when passing through a lens plane.  The
deflection angles may be calculated from the gradient of a lensing potential,
which is connected to the projected matter distribution on the lens planes via
a Poisson equation.

A light ray reaching the observer from a given angular position
$\vect{\theta}$ can then be traced back to the angular position $\vect{\beta}$
of its source at given redshift $z^\mrm{S}$, thereby defining the lensing map
\begin{equation}
\label{eq:def_lensing_map}
\mrm{L}:\;\mbb{P}^\mrm{I}\to\mbb{P}^\mrm{S}:\;\vect{\theta}\mapsto \vect{\beta}
\end{equation}
from the image plane $\mbb{P}^\mrm{I}$ to the source plane $\mbb{P}^\mrm{S}$.
The distortion matrix $\matr{A}=\parder{\vect{\beta}}{\vect{\theta}}$, i.e. the
Jacobian of the map, quantifies the magnification and distortion of images of
small sources induced by gravitational lensing.  The (signed) magnification
$\mu$ of an image is given by the inverse determinant of the distortion
matrix:
\begin{equation}
\mu=\left(\det{\matr{A}}\right)^{-1}.
\end{equation}
The decomposition \citep{ScEhFa92}
\begin{multline}
\label{eq:distortion_decomposition}
\matr{A}=
\begin{pmatrix}
\cos\varphi & \sin\varphi\\
-\sin\varphi&\cos\varphi
\end{pmatrix}
\begin{pmatrix}
1-\kappa-\gamma_1 & -\gamma_2\\
-\gamma_2& 1-\kappa+\gamma_1
\end{pmatrix}
\end{multline}
of the distortion matrix defines the image rotation angle $\varphi$, the
convergence $\kappa$, and the complex shear $\gamma=\gamma_1+\mrm{i}\gamma_2$.  The
reduced shear $g=\gamma/(1-\kappa)$ determines the major-to-minor axis ratio
\begin{equation}
r=\left|\frac{1+|g|}{1-|g|}\right|
\end{equation}
of the elliptical images of sufficiently small circular sources.  The
determinant and trace of the distortion matrix may be used to categorise
images \citep{ScEhFa92}:\footnote{
In \citet{ScEhFa92}, the trace and determinant of the symmetric part [i.e. second factor on the r.h.s. of Eq.\eqref{eq:distortion_decomposition}] of the distortion matrix $\matr{A}$ have been used. However, the determinant and the sign of the trace of $\matr{A}$ and its symmetric part are identical for $|\varphi|<\pi/2$.
}
\begin{itemize}
\item type I: $\det\matr{A}>0$ and $\tr\matr{A}>0$,
\item type II: $\det\matr{A}<0$,
\item type III: $\det\matr{A}>0$ and $\tr\matr{A}<0$.
\end{itemize}
In all situations relevant for this work, images of type II and type III
belong to sources that have multiple images. In the following, we will often
consider type II and III images together as
\begin{itemize}
\item type \IIorIII: $\det\matr{A}<0$ or $\tr\matr{A}<0$.
\end{itemize}

In this paper, we want to study how often one can expect to observe images
with certain lensing properties, e.g. highly magnified or strongly distorted
images.  In order to quantify the frequency of rays with a given property $p$,
we define the optical depth
\begin{equation}
\label{eq:def_tau_I}
\tau^\mrm{I}_p=
\frac{\int_{\mbb{P}^\mrm{I}}\diff[2]{\vect{\theta}}1_p(\vect{\theta})}
{\int_{\mbb{P}^\mrm{I}}\diff[2]{\vect{\theta}}},
\end{equation}
where
\begin{equation*}
1_p(\vect{\theta})=
\begin{cases}
1& \text{if ray}(\vect{\theta})\text{ has property }p, \text{ or}\\
0& \text{otherwise.}
\end{cases}
\end{equation*}
For a uniform distribution of images in the image plane, $\tau^\mrm{I}_p$
estimates the fraction of images of sufficiently small sources that have the
property $p$. Furthermore, we define the optical depth
\begin{equation}
\label{eq:def_tau_S}
\tau^\mrm{S}_p= \frac
{\int_{\mbb{P}^\mrm{I}}\diff[2]{\vect{\theta}}\left|\mu(\vect{\theta})\right|^{-1}
1_p(\vect{\theta})}
{\int_{\mbb{P}^\mrm{I}}\diff[2]{\vect{\theta}}\left|\mu(\vect{\theta})\right|^{-1}}
\end{equation}
to estimate the average fraction of images with properties $p$ for a
uniform distribution of sources in the source plane.  These
optical depths (assumed to be smooth) will be used to define corresponding
probability density functions (pdf) for the magnification:
\begin{equation}
\label{eq:def_pdf_mu}
\pdf^\mrm{I/S}(\mu')=\totder{}{\mu'}\tau^\mrm{I/S}_{\mu(\vect{\theta})\leq \mu'}\;.
\end{equation}

Compared to $\tau^\mrm{I}_p$, the optical depth $\tau^\mrm{S}_p$ takes into
account that areas in the image plane with higher magnification map to smaller
areas in the source plane. This aspect of magnification bias leads to a lower
image density in areas of higher magnification for volume-limited surveys. In
magnitude-limited surveys, however, magnification can push images that would
otherwise be too faint to be observed above the detection threshold. This
aspect of magnification bias counteracts the previous one, but will not be
discussed in this paper since it depends sensitively on the luminosity
distribution of the source population.

Note that $\tau^\mrm{I}_p$ and $\tau^\mrm{S}_p$ differ from the optical depths
\begin{equation}
\label{eq:def_tilde_tau_S}
\tilde{\tau}^\mrm{S}_p=
\frac
{\int_{\mbb{P}^\mrm{S}}\diff[2]{\vect{\beta}} 1_p(\vect{\beta})}
{\int_{\mbb{P}^\mrm{S}}\diff[2]{\vect{\beta}}}.
\end{equation}
discussed, e.g., by \cite{ScEhFa92}, which quantify the fraction of sources
whose images have certain properties. The methods used in this paper do not
yield enough information to deduce $\tilde{\tau}^\mrm{S}_p$ in general. We
will therefore restrict our discussion to $\tau^\mrm{I}_p$ and
$\tau^\mrm{S}_p$.

Roughly speaking, $\tau^\mrm{I}_p$ weights lensing events by their area on the
sky, $\tau^\mrm{S}_p$ weights by the number of images, and
$\tilde{\tau}^\mrm{S}_p$ weights by the number of sources.
In the absence of multiple images, $\tau^\mrm{S}_p$ and $\tilde{\tau}^\mrm{S}_p$  would be identical. For strongly lensed properties such
as considered in this paper, they can be significantly different:
 For $\tau^\mrm{S}_p$, each of the multiple images of a source
contributes individually, whereas all images of the same source contribute a
single event event to $\tilde{\tau}^\mrm{S}_p$.  Consequently, for a given
number $N^S$ of uniformly distributed sources in the source plane,
$N^S\tau^\mrm{S}_p$ is the expected number of images with property $p$,
whereas $N^S\tilde{\tau}^\mrm{S}_p$ gives the expected number of
sources. Obviously, the number of images is easier to count in observations
than the number of sources.

\subsection{The lens planes}
\label{sec:lens_planes}
Our methods for reconstructing the observer's backward light cone, for
splitting it into a series of lens planes, and for calculating the matter
distribution and deflection angles on these planes are generally similar to
those used by \cite{JaSeWh00}. They differ, however, in a number of important
details which reflect our wish to take full advantage of the unprecedented
statistical power offered by the large volume and high spatial and mass
resolution of the Millennium Simulation.  Here, we give a brief outline of our
algorithms, reserving a detailed description for \cite{HilbertEtal07b}.

The Millennium Simulation~\citep{SpringelEtAl05MSReview} is an $N$-body
simulation of cosmological structure formation in a flat $\Lambda$CDM universe
with a matter density of $\Omega_\mrm{M}=0.25$ (in terms of the critical
density), a cosmological constant with $\Omega_\Lambda=0.75$, a Hubble constant
$h=0.73$ in units of $100\,\mrm{km}\mrm{s}^{-1}\Mpc^{-1}$, a primordial
spectral index $n=1$ and a normalisation parameter $\sigma_8=0.9$ for the
linear density power spectrum.  The simulation followed $N\approx10^{10}$
particles of mass $m_\mrm{p}=8.6 \times 10^8 h^{-1}\,\Msolar$ in a cubic region of
comoving side $L=500h^{-1}\,\Mpc$ (assuming periodic boundary conditions) from
redshift $z=127$ to the present using a TreePM version of \mbox{\textsc{gadget}-2} \citep{Springel05GADGET2}. The force softening length was chosen to be
$5h^{-1}\,\kpc$ comoving.  Snapshots of the simulation were stored on disk at 64
output times spaced approximately logarithmically in expansion factor for
$z\ge1$ and at roughly $200\,\Myr$ intervals after $z=1$.

Since the fundamental cube of the simulation is too small to trace rays back
to high redshift in a single replication, we have to make use of the
periodicity to construct our light cones.  In order to reduce the repetition
of structure along long lines-of-sight (LOS) through this lattice-periodic
matter distribution, we chose the LOS to be in the direction
$\vect{n}=(1,3,10)$. This results in a comoving period of $5.24h^{-1}\,\Gpc$ along
the LOS, giving the first image of the origin at $z=3.87$. It also allows us
to maintain periodicity perpendicular to the LOS with a rectangular unit cell
of dimension $1.58h^{-1}\,\Gpc\times1.66h^{-1}\,\Gpc$ comoving. This periodicity allows us
to use Fast-Fourier-Transform (FFT) methods \citep[e.g.,][]{JaTu65_FFT,FrJo05_FFTW3} to obtain the lensing potential and its derivatives on the
lens planes.

To construct the matter distribution in the observer's backward light cone, we
partitioned space into a series of redshift slices, each perpendicular to our
chosen LOS and containing the part of the light cone closer to one of the
snapshot redshifts than to its neighbours. The matter distribution within each
such slice was then approximated by the stored particle data at the time of
the corresponding snapshot, was projected onto the lens plane, and was placed
at the comoving distance corresponding to the snapshot's redshift.  In order
to reduce the shot noise from individual particles, we employed an adaptive
smoothing scheme. Each particle was smeared out into a cloud with projected
surface mass density
\begin{equation}
 \Sigma_\mrm{p}(\vect{x})=
\begin{cases}
\frac{3 m_\mrm{p}}{\pi r_\mrm{p}^2}\left(1-\frac{|\vect{x}-\vect{x}_\mrm{p}|^2}{r_\mrm{p}^2}\right)^{2}, & |\vect{x}-\vect{x}_\mrm{p}|<r_\mrm{p},\\
 0, &  |\vect{x}-\vect{x}_\mrm{p}| \ge r_\mrm{p},
\end{cases}
\end{equation}
where $\vect{x}$ denotes comoving position on the lens plane, $\vect{x}_\mrm{p}$ is
the projected comoving particle position, and $r_\mrm{p}$ denotes the comoving
distance to the 64th nearest neighbour particle in \emph{three} dimensions
(i.e. before projection).

To calculate the lensing potential from the projected matter density, we used
a particle-mesh particle-mesh (PMPM) method.  The effective spatial resolution
of the Millennium Simulation is about $5h^{-1}\,\kpc$ comoving in dense
regions, where the particles' smoothing lengths become comparable to softening length of the simulation, i.e. $r_\mrm{p}\sim5h^{-1}\,\kpc$. Hence, a mesh spacing of $2.5h^{-1}\,\kpc$ comoving is required to exploit the
numerical data without degradation. A single mesh of this spacing covering the
whole periodic area of the lens plane (i.e. $1.58h^{-1}\,\Gpc\times1.66h^{-1}\,\Gpc$ comoving)
would, however, be too demanding both to compute and to store. We therefore
split the lensing potential $\Psi$ into long-range and short-range parts
defined in Fourier space by:
\begin{subequations}
\begin{align}
\Psi_\mrm{long}(\vect{k})&=\Psi(\vect{k})\exp\left(-l_\mrm{split}^2\vect{k}^2\right)\text{, and}\\
\Psi_\mrm{short}(\vect{k})&=\Psi(\vect{k})\left[1-\exp\left(-l_\mrm{split}^2\vect{k}^2\right)\right].
\end{align}
\end{subequations}
The comoving splitting length $l_\mrm{split}=0.175h^{-1}\,\Mpc$ characterises the
spatial scale of the split. We use a $16384\times16384$ mesh covering the whole
periodic area of the lens plane to calculate $\Psi_\mrm{long}$ from the projected surface mass density by an FFT~\citep{FrJo05_FFTW3} method.
The long-range potential is calculated once and then stored on disk for each of the lens planes. To calculate $\Psi_\mrm{short}$, a fine
mesh with $2.5h^{-1}\,\kpc$ spacing is used. This mesh only need cover a
relatively small area around regions where the potential is required,
i.e. where light rays intersect the lens plane. Because of the short range of
$\Psi_\mrm{short}$, periodic boundary conditions can be used on the fine mesh,
provided points close to its boundary are excluded from subsequent
analysis. Therefore, FFT methods can used without `zero padding' to calculate $\Psi_\mrm{short}$ on the fine mesh.
The long- and short-range contributions to the deflection angles and
shear matrices (i.e. the second derivatives of the lensing potential) are
calculated on the two meshes by finite differencing of the potentials. The
values between mesh points are obtained by bilinear interpolation. The
resulting deflection angles and shear matrices at ray positions can then be
used to advance the rays and their associated distortion matrices from one
plane to the next.

\subsection{Ray sampling}
\label{sec:ray_sampling}
In order to estimate optical depths and magnification distributions, we shoot
random rays through our set of lens planes.  In doing this, we neglect
correlations between the matter distributions on different lens planes. This
allows us to pick random points on each lens plane as we propagate rays back in
time, significantly simplifying and accelerating our code. Since the comoving
separation between the lens planes is large ($\sim 100 h^{-1}\,\Mpc$), and the shear
matrices are dominated by small-scale structure ($\lesssim 1 h^{-1}\,\Mpc$), this
assumption is very well justified for the purposes of the current paper.

On each lens plane, 40 fields of about $40h^{-1}\,\Mpc\times40h^{-1}\,\Mpc$ comoving were
selected at random. Within each of these fields, the shear matrix was
calculated at 16 million random positions by our PMPM algorithm. The resulting
$6.4\times 10^8$ shear matrices for each plane were then stored on disk
together with the positions at which each had been computed.

The shear matrices from all lens planes were next combined at random to
produce 640 million simulated LOS from the observer back to high
redshift. Imagining a small circular source at the position where each of
these LOS intersects a particular lens plane, we can combine the shear
matrices of the ray from all lower redshift planes to obtain the trace
$\tr\matr{A}$ of the distortion matrix, the magnification
$\mu=(\det\matr{A})^{-1}$, and the length-to-width ratio $r$ of the source's
image. The measured fractions of rays with certain properties, e.g. a large
magnification, can then be used to estimate the corresponding optical depth to
the redshift of the chosen plane.

This sampling method assumes that the rays are uniformly distributed in the
image plane. The observed number $N_p$ of rays with a particular property $p$,
when compared to the total number of rays $N$, is thus a straightforward
Monte-Carlo estimate (without importance sampling) of the optical depth
$\tau^\mrm{I}_p$:
\bse
\begin{equation}
\tau^\mrm{I}_p\approx\frac{N_p}{N}\;.
\end{equation}
The calculation of the corresponding optical depth $\tau^\mrm{S}_p$
requires using the individual magnifications $\mu(i)$ of the rays
$i=1,\ldots,N$ as statistical weights:
\begin{equation}
\tau^\mrm{S}_p\approx\frac{\sum_{i=1}^N \left|\mu^{-1}(i)\right|1_p(i)}{\sum_{i=1}^N \left|\mu^{-1}(i)\right|}\;.
\end{equation}
\ese

The matter distribution on our lens planes is guaranteed to be periodic,
smooth and non-singular as a result of the adaptive smoothing we
use. Furthermore, a large and random area of the image plane is mapped onto an
equally large area in the source plane by our lensing
map~\eqref{eq:def_lensing_map}. Thus a representative ray sample should
satisfy [see Eq.~\eqref{eq:mu_sample_sum} in Appendix~\ref{sec:integrals_of_mu}]:
\begin{equation}
 1\approx\frac{1}{N}\sum_{i=1}^N\mu^{-1}(i)\;.
\end{equation}
For our ray sample, we find this relation to be satisfied to quite high
precision; for all source planes the deviation is smaller than $0.003$.

Our ray sampling technique neglects correlations between the structure on
different lens planes. The effects of the lens environment on scales smaller
than $100 h^{-1}\,\Mpc$ comoving should be correctly represented, however, and the
effects of uncorrelated fluctuations in the density of foreground and
background matter are also included properly. This simple procedure should
thus give accurate results in the context where we use it, but we note that it
does not allow the construction of extended images of extended sources. This
can be done by relatively straightforward extensions of our methods which we reserve for
future papers.

\section{Results}
\label{sec:results}
\subsection{The magnification distribution}
\label{sec:magnification_distribution}

\begin{figure}
\centerline{\includegraphics[width=1\linewidth]{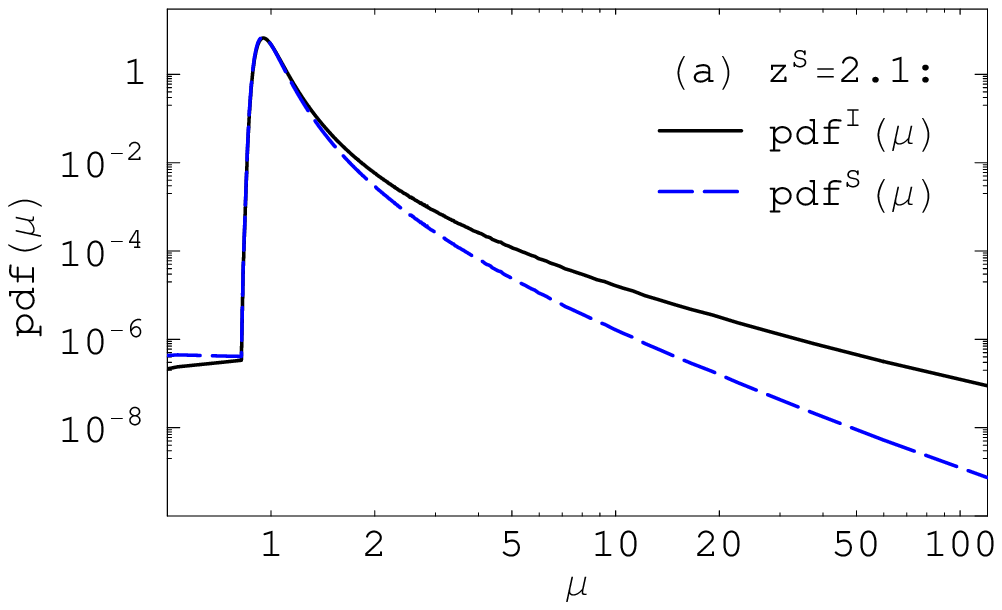}}
\centerline{\includegraphics[width=1\linewidth]{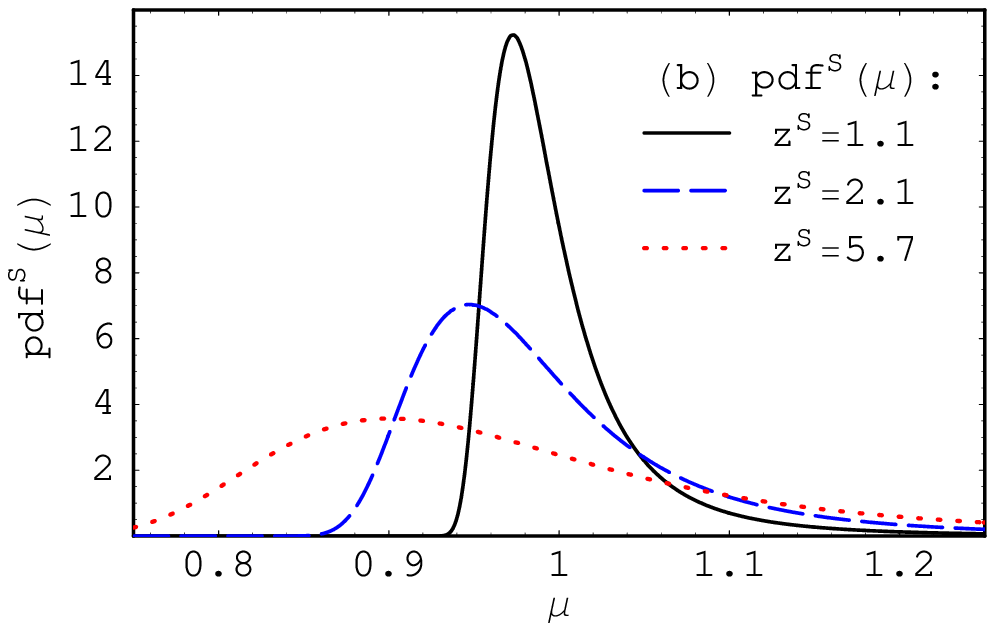}}
\caption{
\label{fig:pdf_of_mu}
The probability density \mbox{$\pdf(\mu)$} of the magnification $\mu$. (a)
Logarithmic plot comparing \mbox{$\pdf^\mrm{I}(\mu)$} (solid line) and
\mbox{$\pdf^\mrm{S}(\mu)$} (dashed lines) for a source redshift of
\mbox{$\zS=2.1$}. (b) Linear plot comparing the \mbox{$\pdf^\mrm{I}(\mu)$}
around $\mu=1$ for different source redshifts, \mbox{$\zS=1.1$} (solid line),
\mbox{$\zS=2.1$} (dashed lines) and \mbox{$\zS=5.7$} (dotted lines).  }
\end{figure}

\begin{figure}
\centerline{\includegraphics[width=1\linewidth]{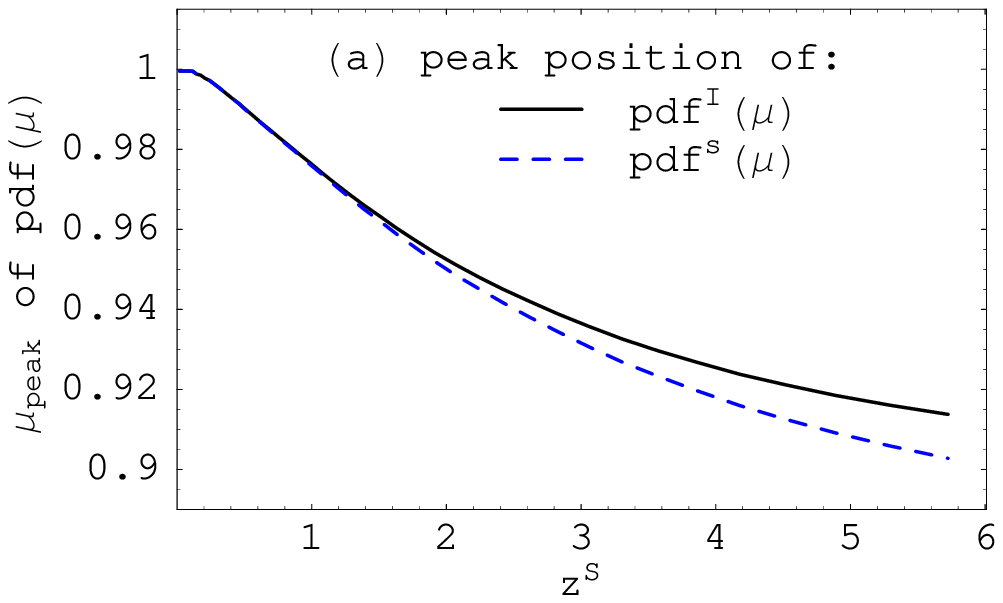}}
\centerline{\includegraphics[width=1\linewidth]{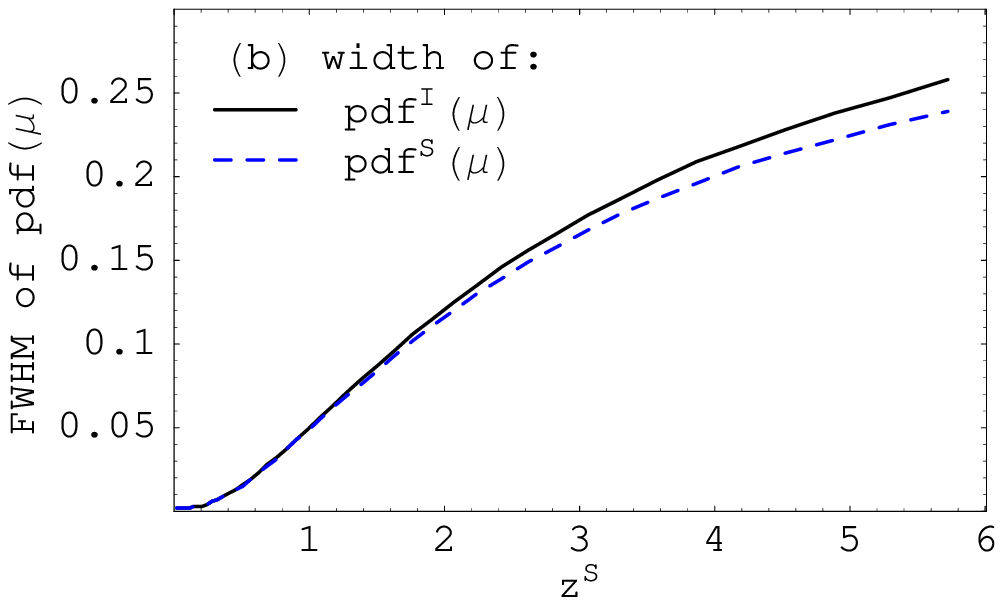}}
\caption{
\label{fig:mode_and_width_of_pdf_of_mu}
The peak position (a) and the width (b) of the probability densities
\mbox{$\pdf^\mrm{I}(\mu)$} (solid lines) and \mbox{$\pdf^\mrm{S}(\mu)$} (dashed
lines) for the magnification $\mu$ as a function of source redshift $\zS$. The
distributions get broader and more skew with increasing $\zS$. }
\end{figure}

From the magnifications of our random rays, we have estimated the probability
density functions $\pdf^\mrm{I}(\mu)$ and $\pdf^\mrm{S}(\mu)$.  These are
compared for a source redshift \mbox{$\zS=2.1$} in
Fig.~\ref{fig:pdf_of_mu}a. One can readily see the stronger fall-off to high
magnification for $\pdf^\mrm{S}(\mu)$, which is a consequence of the relation $\pdf^\mrm{S}(\mu)\approx|\mu|^{-1}\pdf^\mrm{I}(\mu)$ [see
Eq.~\eqref{eq:pdf_mu_relations} in the Appendix]. The asymptotic behaviour predicted from
catastrophe theory \citep{ScEhFa92}, i.e. $\pdf^\mrm{I}(\mu)\propto\mu^{-2}$
and $\pdf^\mrm{S}(\mu)\propto \mu^{-3}$, is reached for magnifications $\mu
\gtrsim 20$.

Probability density functions $\pdf^\mrm{S}(\mu)$ for different source
redshifts $\zS$ are shown in Fig.~\ref{fig:pdf_of_mu}b.  With increasing
$\zS$, the peak of the distribution broadens and moves to lower $\mu$, whereas
the high-$\mu$ tail increases in amplitude. The peak positions and the widths of the
pdfs, i.e. their modes $\mu_\mrm{peak}$ and full-widths-at-half-maximum FWHM, are plotted as functions of source redshift in
Fig.~\ref{fig:mode_and_width_of_pdf_of_mu}.  The shift of the peak with
increasing redshift balances the heavier tail so that $\int \mu
\pdf^\mrm{S}(\mu)\diff\mu\approx 1$ for all redshifts [see
Eq.~\eqref{eq:pdf_mu_integrals}].  These results for peak
position and width are in good agreement with those of \cite{Valageas00} and
\cite{FlukeEtal02}, who considered similar cosmologies.

There is a lower bound to the magnification of images of type I, which is realized for rays which propagate through empty cones (i.e., for which the matter density vanishes along their path) and which are subject to no shear effects \citep[e.g.,][]{DyRo72,SeSc92}. No light ray can diverge more strongly than such an empty-beam ray. Lower magnifications can only be produced for overfocussed rays which then belong to Type II or III.
A simple way of calculating this lower bound $\mu_\mrm{min}$ for a flat universe is given in Appendix~\ref{sec:empty_beam_magnification}.
The steep rise in the probability density of the magnification at $\mu\approx0.83$ seen in Fig.~\ref{fig:pdf_of_mu}(a) is substantially larger than
the theoretical bound $\mu_\mrm{min}=0.69$ for $\zS=2.1$, indicating that there are no real empty cones in a realistic universe, which is in agreement with findings by \cite{VaWh03}.

\subsection{Strong-lensing optical depths}
\label{sec:optical_depths}

\begin{figure}
\centerline{\includegraphics[width=1\linewidth]{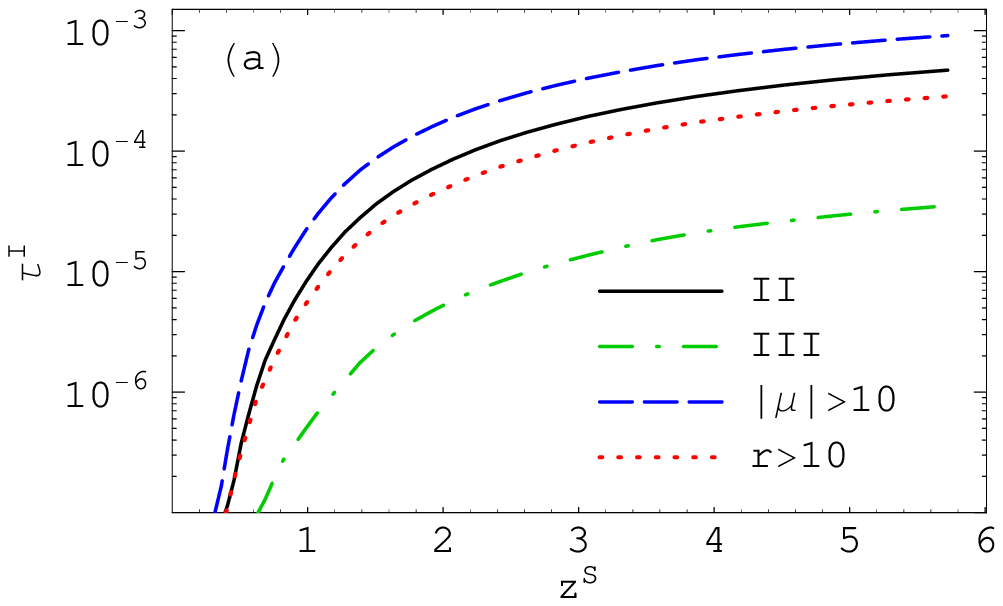}}
\centerline{\includegraphics[width=1\linewidth]{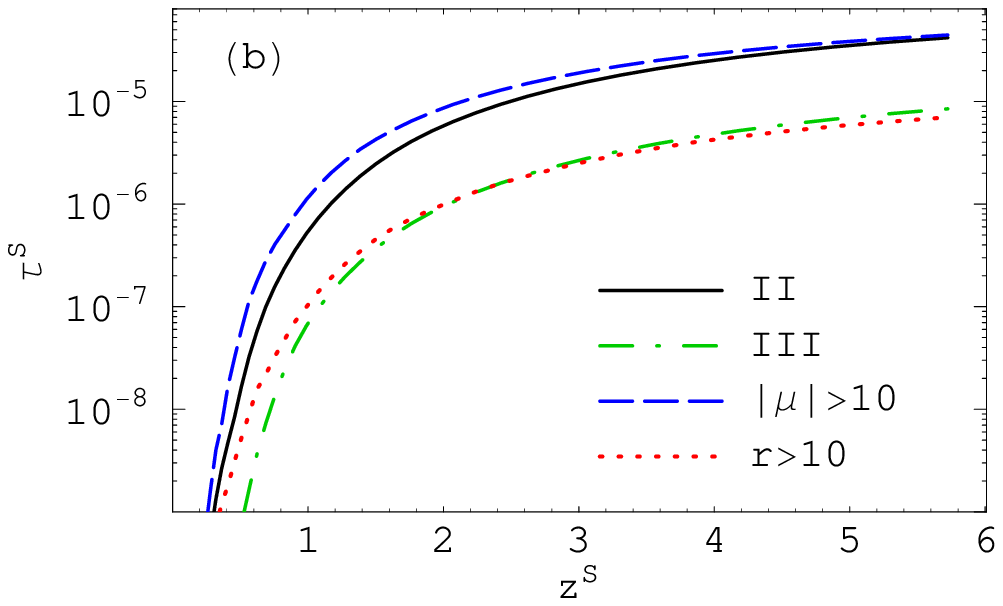}}
\caption{
\label{fig:tau}
Optical depths for images of small circular sources of type II (solid lines),
of type III (dash-dotted lines), with large magnification (dashed lines) and with
large length-to-width ratio (dotted lines), assuming a uniform
distribution of images in the image plane (a), and a uniform distribution of
sources in the source plane (b). Note that the optical depths are
significantly smaller in the latter case, and that the relative optical
depths for different types of strong lensing are not the same in the
two cases. }
\end{figure}

Sufficiently concentrated matter clumps between distant light sources and the
observer can give rise to highly magnified, strongly distorted, or multiple
images. We refer to such phenomena as strong lensing. In order to quantify the
amount of strong lensing expected in a $\Lambda$CDM universe with the
parameters of the Millennium Simulation, we used our large set of random rays
to estimate:
\begin{itemize}
\item the fraction with \mbox{$\det\matr{A}<0$} \mbox{(type II)},
\item the fraction with \mbox{$\det\matr{A}>0$} and \mbox{$\tr\matr{A}<0$}
\mbox{(type III)},
\item the fraction with \mbox{$\det\matr{A}<0$} or \mbox{$\tr\matr{A}<0$},
i.e. the sum of the two previous classes (type \IIorIII),
\item the fraction with a length-to-width ratio \mbox{$r>10$} for images of
sufficiently small circular sources, and
\item the fraction with magnification \mbox{$|\mu|>10$}.
\end{itemize}
The corresponding optical depths $\tau_p^\mrm{I}(\zS)$ and
$\tau_p^\mrm{S}(\zS)$ are plotted in Fig.~\ref{fig:tau} as functions of the
source redshift $\zS$. Since all the image properties we consider are (either by definition or at least statistically)
associated with large magnifications, the optical depths $\tau_p^\mrm{S}(\zS)$
are always a factor 5 to 20 smaller than the corresponding
$\tau_p^\mrm{I}(\zS)$.

The optical depths for $r>10$ are 2 to 20 times smaller than those for
$|\mu|>10$. Similar results have been found by \cite{DaHoHe04} and by
\cite{LiEtal05}. Evidently, the optical depth for highly magnified images does not provide a reliable estimate for the probability of images with a large length-to-width ratio.

The optical depth $\tau^{\mrm{S}}_{r>10}$ may be a reasonable
approximation to the optical depth for giant arcs with length-to-width
ratio $r>10$, since both finite source size and finite source ellipticity
affect this particular property only weakly. Moreover the two effects work in
opposite directions.  For example, \cite{LiEtal05} found that the optical depth for 
${r>10}$ is almost identical to the optical depth for arc images with length-to-width-ratios $>10$ of elliptical sources with an effective diameter of $1\,\arcsecs$. 

All our optical depths show a strong dependence on source redshift, similar to
that previously noted by \cite{WaBoOs04} and \cite{LiEtal05}. Comparing to
their results in detail, however, we find a somewhat stronger redshift
dependence, resulting in 2 to 3 times higher optical depths at $\zS>0.5$.
There are various possible explanations for this discrepancy. The simulations
they use have 3 to 20 times worse mass resolution than the Millennium
Simulation. In addition, \cite{WaBoOs04} reduced their spatial resolution on lens
planes with increasing redshift, thereby potentially missing some low mass
lenses, whereas we always use the full spatial resolution permitted by the
force resolution of the Millennium Simulation.
Furthermore, \cite{WaBoOs04} measured the fraction of multiply
imaged sources for which at least one image has $|\mu|>10$. Thus they do not
include sources with a single highly magnified image, as can occur for a
marginally subcritical lens. Moreover, they do not account for the fact
that multiply imaged sources can give rise to more than one image with
$|\mu|>10$. Similarly, \cite{LiEtal05} only considered the magnification $\mu$ and length-to-width-ratio $r$ of the brightest image of multiply imaged sources for the calculation of the optical depths for ${|\mu|>10}$ and ${r>10}$, although they took into account all giant-arc images for a given population of extended sources in their calculation of the optical depth for giant arcs.
Furthermore, they only considered massive clusters and neglected possible contributions from foreground and background matter.
These decisions caused them to underestimate the total cross sections for strong lensing (at least for smaller sources).

Fold singularities generically produce strongly magnified and distorted images as pairs, and sources near cusps may give rise to one or three strongly magnified or distorted images \citep{ScEhFa92}. Therefore, sources with multiple highly magnified or strongly distorted images might explain a substantial part of the discrepancy between our results and those of \cite{WaBoOs04} and \cite{LiEtal05}.
With our methods, we cannot quantify this effect, but we discuss the other effects further in the following sections.

\subsection{Lens properties}
\label{sec:lens_properties}

\begin{figure}
\centerline{\includegraphics[width=1\linewidth]{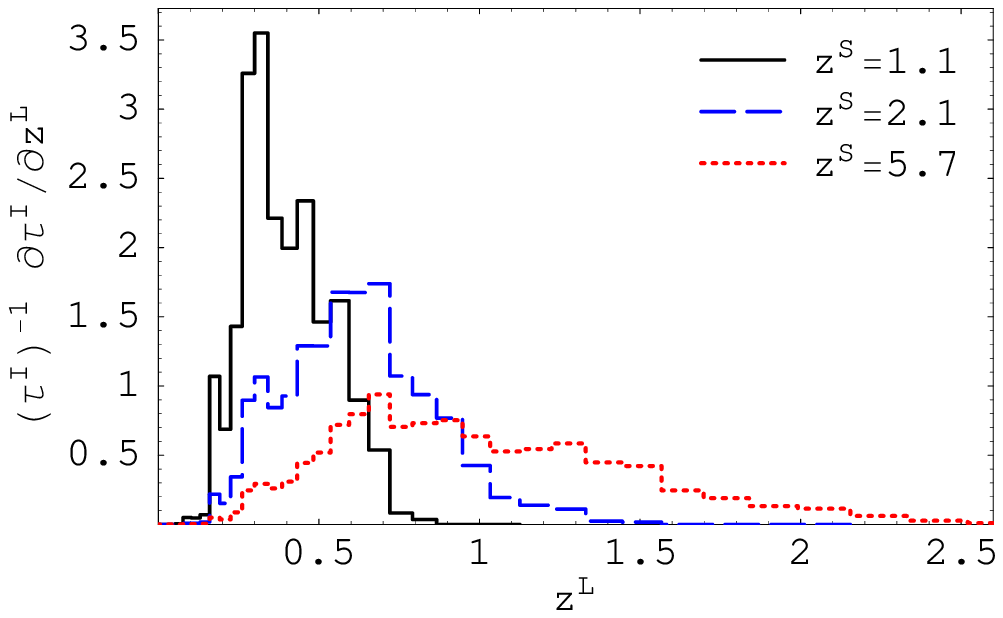}}
\caption{
\label{fig:sigma_of_zL}
The cross section \mbox{$\partial\tau^\mrm{I}/\partial z^\mrm{L}$} for rays of
type \IIorIII\ as a function of lens redshift $z^\mrm{L}$ for sources at
redshift \mbox{$z^\mrm{S}=1.1$} (solid line), $z^\mrm{S}=2.1$ (dashed line),
and \mbox{$z^\mrm{S}=5.7$} (dotted line). Even for high redshift sources, the
typical lens redshift is relatively low. }
\end{figure}

\begin{figure}
\centerline{\includegraphics[width=1\linewidth]{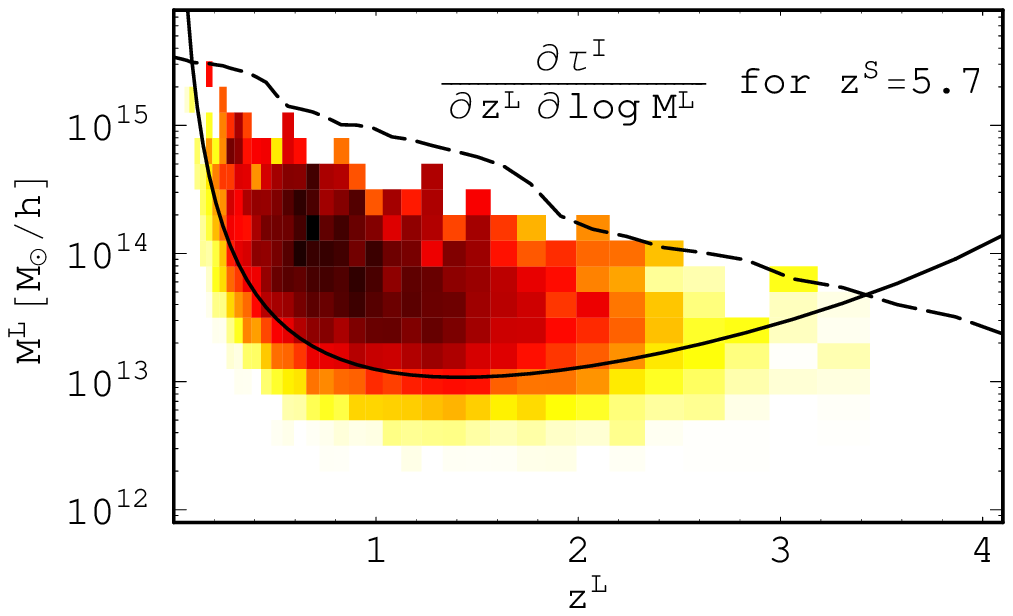}}
\caption{
\label{fig:sigma_of_zL_and_LogML}
The cross section \mbox{$\partial\tau^\mrm{I}/\partial \zL \partial \log M^\mrm{L}$}
for rays of type \mbox{\IIorIII} and sources at redshift \mbox{$z^\mrm{S}=5.7$}
as a function of the redshift $\zL$ and halo mass $M^\mrm{L}$ of the lens.
Darker areas correspond to higher cross sections (on a logarithmic scale).
The dashed line marks the mass of the largest DM halo in
the Millennium Simulation at each redshift. The solid line joins masses for
which a spherical NFW halo of typical concentration would have Einstein radius
$5h^{-1}\,\kpc$, approximately the resolution limit on the lens planes. The
cross section for strong lensing in our ray sample is almost entirely
contained between these two limits.}
\end{figure}

In most cases, the properties of strongly lensed rays, i.e. rays with
$\det\matr{A}<0$, $\tr\matr{A}<0$, $|\mu|>10$, or $r>10$, are predominantly
caused by a single matter clump along the line of sight. We refer to this as
the lens of the ray. In order to find these clumps and to study their
properties, we determined for each strongly lensed ray those lens planes which
were sufficient to produce the relevant property in the single-plane
approximation. Only $2\times10^{-4}$ of all rays had more than one
`sufficient' plane, and we will simply ignore these rays in the following. On
the other hand, there was no sufficient plane for up to 41 percent of the
rays, depending on source redshift and the property considered, so we will
discuss these cases in more detail in Sec~\ref{sec:los_matter}. For most
strongly lensed rays, however, this simple criterion identifies exactly one
lens plane. The redshift of this plane was then taken as the lens redshift
$\zL$ for the ray. For rays of type \IIorIII\ (i.e. rays with $\det\matr{A}<0$
or $\tr\matr{A}<0$), the resulting lens redshift distribution is illustrated in
Fig.~\ref{fig:sigma_of_zL}. Here we plot the cross section
$\partial\tau^\mrm{I}/\partial \zL$ as a function of lens redshift $\zL$ for
various source redshifts $\zS$. Even for sources at redshift $\zS=5.7$, most
of the lenses have redshift $\zL<2.5$. The relatively low cross section at
higher $\zL$ reflects both the lower abundance of massive halos and the less
favourable geometry for lensing at these redshifts (see
Fig.~\ref{fig:sigma_of_zL_and_LogML}). The lens redshift distributions for
rays with $|\mu|>10$ and with $r>10$ are almost indistinguishable from that of
Fig.~\ref{fig:sigma_of_zL} despite the different total optical depths.
 
We studied not only the redshift of the clumps acting as strong
lenses, but also their masses. All significant matter concentrations have
already been identified as DM halos in the simulation and their masses and
central positions are available in the simulation archive. We first projected
the centres of all halos onto the lens planes in the same way as was done for
the particles. For each strongly lensed ray and for all DM halos close to the
point where the ray intersects a lens plane, we determined the ratio
$M^\mrm{L}/b$, where $M^\mrm{L}$ is the conventional halo mass (defined as the
mass within a sphere with mean enclosed density 200 times the cosmological
mean), and $b$ denotes the impact parameter of the ray with respect to halo
centre.

\begin{figure}
\centerline{\includegraphics[width=1\linewidth]{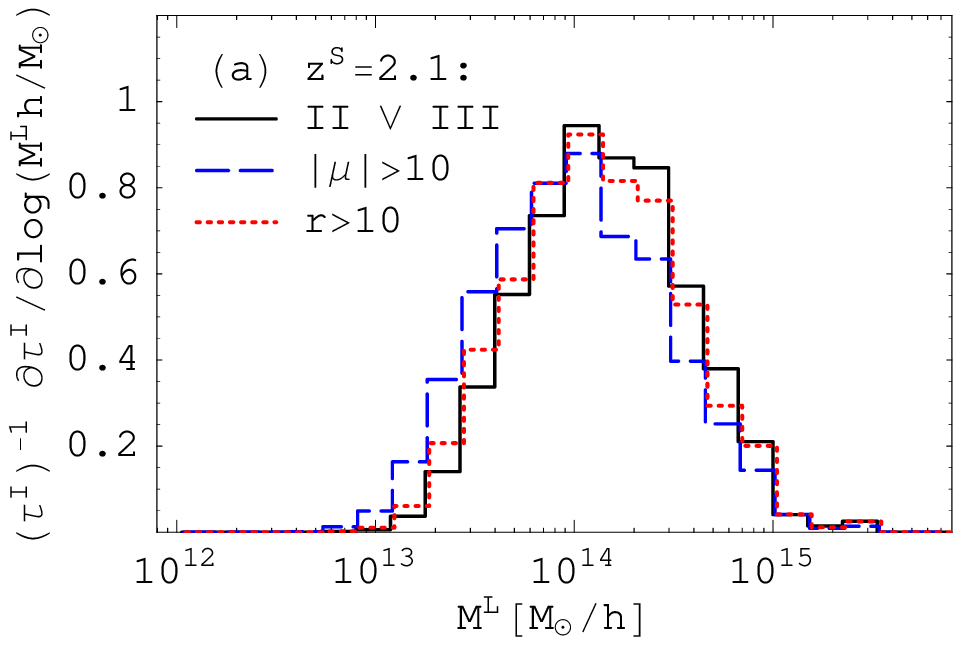}}
\centerline{\includegraphics[width=1\linewidth]{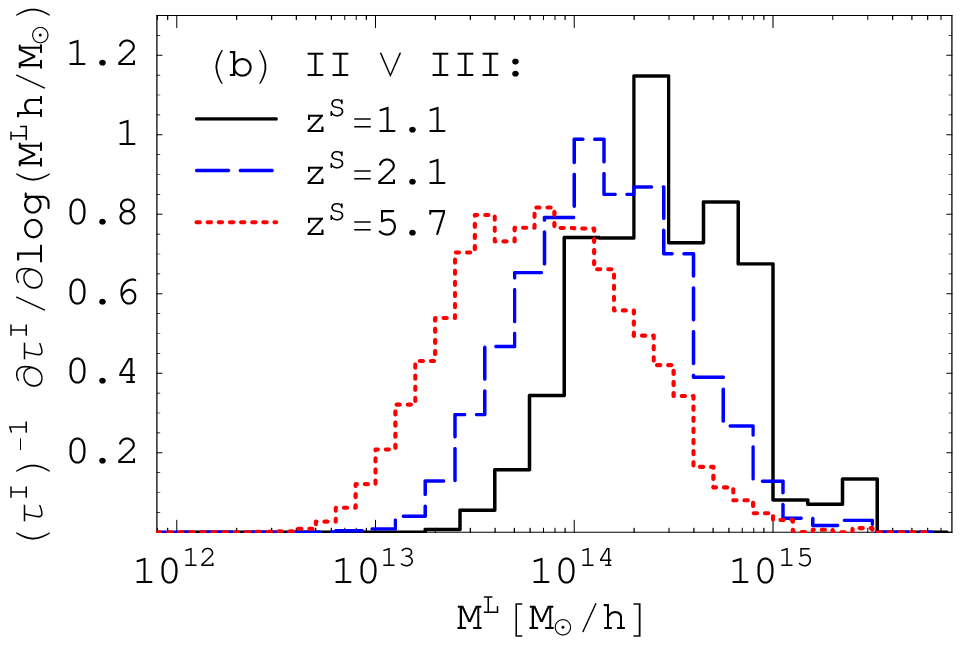}}
\caption{
\label{fig:sigma_of_logML}
The cross section \mbox{$\partial\tau^\mrm{I}/\partial \log M^\mrm{L}$} as a
function of the mass of the lensing halo $M^\mrm{L}$ (see text). Panel (a)
compares rays of type \IIorIII\ (solid line), rays with \mbox{$|\mu|>10$}
(dashed line), and rays with \mbox{$|r|>10$} (dotted line) for source redshift
\mbox{$z^\mrm{S}=2.1$}. (b) compares rays of type \IIorIII\ for sources at
redshift \mbox{$z^\mrm{S}=1.1$} (solid line), $z^\mrm{S}=2.1$ (dashed line),
and \mbox{$z^\mrm{S}=5.7$} (dotted line). The lens mass distribution is almost
independent of the type of strong lensing event, but it shifts towards
lower masses for higher source redshift.}
\end{figure}

The DM halo with the largest $M^\mrm{L}/b$ on the sufficient plane was then
defined to be the lens of ray.  We discarded from further analysis those 3 to
6 percent of rays for which the largest value of $M^\mrm{L}/b$ was not at
least ten times the second largest value. This cut removed all cases where one
could not easily separate the influence of several neighbouring DM halos, for
example in merging clusters.\footnote{The choice $M^\mrm{L}/b$ is somewhat
arbitrary. We also tried $M^\mrm{L}/b^2$, but a different halo was chosen
only in a few cases, all of which were removed by our imposed ratio cut.}
The resulting distributions of lens masses for rays of type \IIorIII, with
$|\mu|>10$, and with $r>10$ are compared in Fig.~\ref{fig:sigma_of_logML}a,
where the cross sections $\partial\tau^\mrm{I}/\partial \log M^\mrm{L}$ are
plotted for $\zS=2.1$ as a function of lens halo mass $M^\mrm{L}$. Although
the corresponding total optical depths are quite different, their lens mass
distributions are very similar. There is only a small shift toward lower
masses for rays with $|\mu|>10$, and $r>10$ compared to rays of type
\IIorIII. In the following we will restrict discussion to the
latter for simplicity.

In Fig.~\ref{fig:sigma_of_logML}b, the cross section
$\partial\tau^\mrm{I}/\partial \log M^\mrm{L}$ is plotted as a function of
mass for type-\IIorIII\ rays and for various source redshifts $\zS$. The
measured cross section \mbox{$\partial\tau^\mrm{I}/\partial \log M^\mrm{L}$}
vanishes for masses below \mbox{$4\times10^{12}h^{-1}\,\Msolar$} and above
\mbox{$4\times10^{15}h^{-1}\,\Msolar$}.  The main contribution to the optical depth
$\tau^\mrm{I}$ comes from halos with \mbox{$10^{13}h^{-1}\,\Msolar\lesssim
M^\mrm{L}\lesssim 10^{15}h^{-1}\,\Msolar$}.  For higher source redshifts, the cross
section has more weight at lower masses.

The upper mass limit for the cross section simply reflects the fact that there
are no halos more massive than \mbox{$4\times10^{15}h^{-1}\,\Msolar$} in the
simulation. However, the cross section decreases rapidly with increasing
$M^\mrm{L}$ already for \mbox{$M^\mrm{L}>10^{15}h^{-1}\,\Msolar$}. The exponential
decrease in halo abundance with increasing mass apparently dominates over the
increasing cross section of individual halos.

At all source redshifts there is a significant contribution from halos with
\mbox{$M^\mrm{L}<10^{14}h^{-1}\,\Msolar$}. For the small sources considered here,
the set of DM halos causing strong lensing extends to substantially lower
masses than \cite{LiEtal05} and \cite{DaHoHe04} suggest for halos producing
giant arcs. The Millennium Simulation has 12 to 20 times better mass
resolution than the simulations used by these authors. This provides a
considerably better representation of the central halo regions which produce
strong lensing. In addition, halos are inefficient in generating strongly
distorted images for sources with angular extent comparable to or larger than
their Einstein radii.  Thus, neglecting or incorrectly treating halos below a
given mass (e.g. because of limited mass resolution) has a larger effect on
the optical depths for small sources than on those for extended sources --
especially for high source redshifts. Together these effects may explain why
we find 2 to 3 times larger optical depths at high redshift than the values
given by \cite{LiEtal05}.

What sets the lower mass limit for the cross section
\mbox{$\partial\tau^\mrm{I}/\partial \log M^\mrm{L}$}?  The nominal mass
resolution for {\it identifying} DM halos in the Millennium Simulation is
about \mbox{$10^{10}h^{-1}\,\Msolar$}, so there are plenty of halos with
\mbox{$M^\mrm{L}<4\times10^{12}h^{-1}\,\Msolar$}. The identification limit cannot,
in itself, explain the lack of halos less massive than
\mbox{$4\times10^{12}h^{-1}\,\Msolar$} in our sample. However, the regions capable
of causing strong lensing are very small for low-mass halos, so our cross
section estimates may be limited by the resolution of our lens planes;
critical regions with a diameter below the mesh spacing or the effective
gravitational smoothing scale are not resolved.  In order to estimate the mass
limit induced by these effects, we considered spherical NFW halos
\citep{NaFrWh97} with concentration parameter
\begin{equation} \label{eq:c_fit} 
c(M^\mrm{L},\zL)=\frac{9.59}{1+\zL}
\left(\frac{M^\mrm{L}}{10^{14} h^{-1}\,\Msolar}\right)^{-0.102} 
\end{equation}
determined by halo mass and redshift \citep{DolagEtal04}. For given lens and
source redshifts, there is a minimum lens mass for which the Einstein radius
exceeds the resolution limit of our lens plane. We take the latter to be
$5\,\kpc$ comoving, thus requiring a minimum of four mesh points across the
Einstein diameter. This limit takes into account not just the limit imposed by the mesh spacing of $2.5h^{-1}\,\kpc$ ($\sim\!1.7h^{-1}\,\kpc$ in radius), but also the force softening and the smoothing for particles in the halo cores.
The solid line in Fig.~\ref{fig:sigma_of_zL_and_LogML}
shows the resulting minimum mass $M^\mrm{L}$ as a function of $\zL$ for
$\zS=5.7$.  The shading in this plot gives the cross section
\mbox{$\partial^2\tau^\mrm{I}/\partial \zL\partial \log M^\mrm{L}$}. The
region of the \mbox{$(\zL,\log M^\mrm{L})$}-plane with non-zero cross section
is bounded above by the largest halo mass at each redshift (the dashed
line). The lower boundary of this region lies slightly below our analytic
estimate of the resolution limit (the solid line). About 6 percent of the cross section is below the analytic estimate.
Some deviation is expected because this
estimate does not include the effects of intrinsic ellipticity, asymmetries and substructure,  and of the
scatter in concentration of halos of given mass. It also neglects scatter due
to additional matter inhomogeneities along the LOS. These effects should
result in strong lensing by somewhat lower mass halos than our simple
spherical model would indicate~\citep{MeneghettiEtal07,HennawiEtal07,FedeliEtal07ArXiv}. In particular, a high-concentration, prolate
halo with its major axis along the LOS has a greatly enhanced cross section
relative to a spherical halo of the same mass with a typical concentration. Moreover, ellipticity and scatter in concentration lead
to larger cross sections \emph{on average} compared to spherical NFW halos of
the mean concentration.

Besides the resolution limit due to the mesh spacing and local smoothing scale on the lens plane, there is another factor limiting the resolution:
The critical regions of the
relevant halos may not have been simulated to the accuracy required to get fully
converged results in the face of discreteness effects resulting
from the relatively small number of particles in these
regions. According to the criteria given by \cite{PowerEtal03}, only the most
massive halos in the Millennium Simulation at $0.3<z<2$ should have
spherically averaged density profiles which are fully converged at radii
comparable to their Einstein radius. At lower masses and at other redshifts
the particle number in the inner regions is below the value advocated by
\cite{PowerEtal03}. (In contrast, the softening length of the Millennium
Simulation appears adequate to avoid major problems.) When the particle number
is too small, \cite{PowerEtal03} show that simulations typically underestimate
the central concentration of a halo, implying a reduction in its strong
lensing cross section. This effect is a relatively slow function of particle
mass. In addition, strong lensing depends on the projected density
distribution rather than the 3-D density profile which \cite{PowerEtal03}
studied; there are typically two to three orders of magnitude more particles
projected within a halo's Einstein radius than there are within a central
sphere of radius $r_E$. Thus it is unclear how seriously the under-resolution
of halo cores will affect the cross sections we calculate.

\begin{figure}
\centerline{\includegraphics[width=1\linewidth]{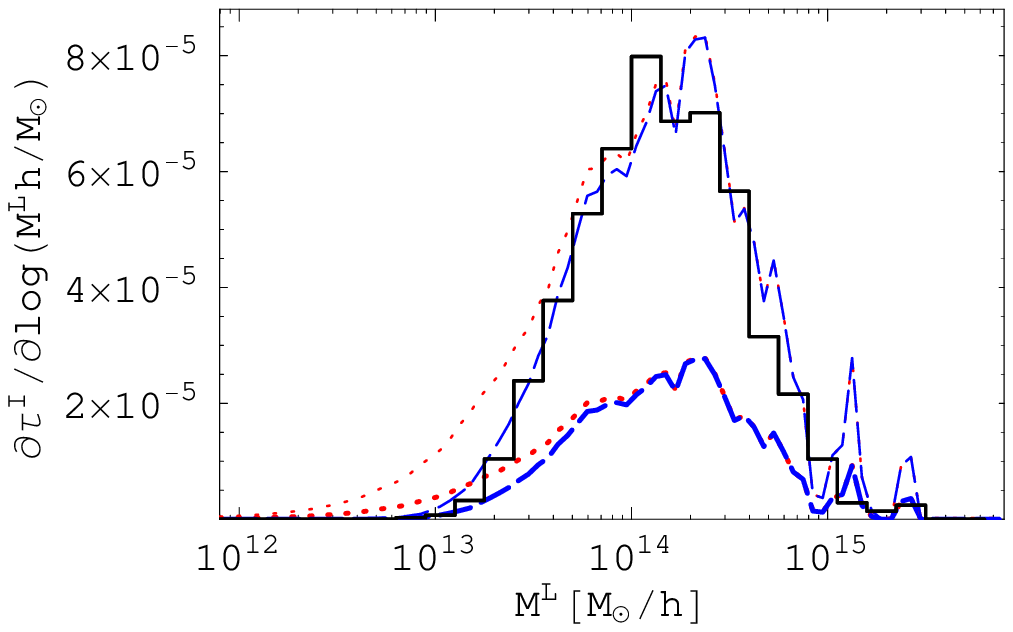}}
\caption{
\label{fig:sigma_of_LogML_NFW}
The cross section \mbox{$\partial\tau^\mrm{I}/\partial \log M^\mrm{L}$} for
rays of type \IIorIII\ to sources at $\zS=2.1$ as a function of the lensing
halo mass $M^\mrm{L}$. The solid histogram gives the direct estimate from
the Millennium Simulation. For all other curves we have replaced each
halo in the simulation by a spherical NFW halo with the same virial mass and 
maximal circular velocity. The heavy dotted line gives the result for
all halos, while the heavy dashed line shows the effect of excluding all halos
with Einstein radius $\rE<5h^{-1}\,\kpc$ comoving.
Thin lines show the result of
scaling these two curves up by a factor of 3, so that their shape can be
compared more easily to the direct estimate from the simulation. The cut-off
in cross section at low halo mass in the simulation appears to correspond well
to that induced by this simple model for the resolution limit. This suggests
that resolution effects reduce our total cross sections by of order 15\%.}
\end{figure}

To obtain a rough estimate of how much optical depth we lose due to resolution
effects, we calculated the cross section
\mbox{$\partial\tau^\mrm{I}_{\IIorIII}/\partial \log M^\mrm{L}$} approximating
all DM halos in the Millennium Simulation by spherical NFW halos while either
(i) taking into account or (ii) disregarding halos with an Einstein radius
$\rE<5h^{-1}\,\kpc$ comoving. In doing this, we used the measured maximal circular
velocity of each halo to estimate its concentration parameter, rather than
assuming the concentration to be given by Eq.~\eqref{eq:c_fit}.
The cross sections obtained for the two cases are shown in
Fig.~\ref{fig:sigma_of_LogML_NFW}. 
Due to the scatter in the halo concentrations, the analytical estimate excluding halos with Einstein radius $r_E<5h^{-1}\,\kpc$ extends to lower masses than the limit calculated above by assuming Eq.~\eqref{eq:c_fit} for all halos.
The estimate for
\mbox{$\tau^\mrm{I}_{\IIorIII}$} including halos with Einstein radius $r_E
<5h^{-1}\,\kpc$ is only about 15 percent larger than the estimate excluding such
halos. Moreover, there is no significant contribution to the full estimate
from halos below a few times $10^{12}h^{-1}\,\Msolar$. Indeed, a detailed analysis
shows that the strong-lensing cross section of spherical NFW halos with
$M^\mrm{L}\leq10^{13}h^{-1}\,\Msolar$ decreases exponentially with decreasing mass,
and is not compensated by the increasing number of
halos.

Fig.~\ref{fig:sigma_of_LogML_NFW} also shows the two spherical
halo-based estimates scaled up by a factor of 3. The curve neglecting halos
with small Einstein radii is then a good match to the histogram derived
directly from the simulation.  Hence, the cross section
$\partial\tau^\mrm{I}_{\IIorIII}/\partial \log M^\mrm{L}$ is about three times
as large for simulated halos as for spherical NFW halos, at least for halo
masses $M^\mrm{L}>4\times10^{13}h^{-1}\,\Msolar$ where the resolution limit is
unimportant.  If this result applies also at lower mass, it implies that
limited resolution does not lead us to underestimate total optical depths
substantially, perhaps only by about 15\%.  Note that the missing cross
section corresponds to very small image splittings
$\Delta\theta^\mrm{I}\lesssim 1\,\arcsecs$, a scale where the gravitational effects of
the baryons in the central galaxy are expected to be important.

\subsection{Effects of additional matter along the line of sight}
\label{sec:los_matter}
\begin{figure}
\centerline{\includegraphics[width=1\linewidth]{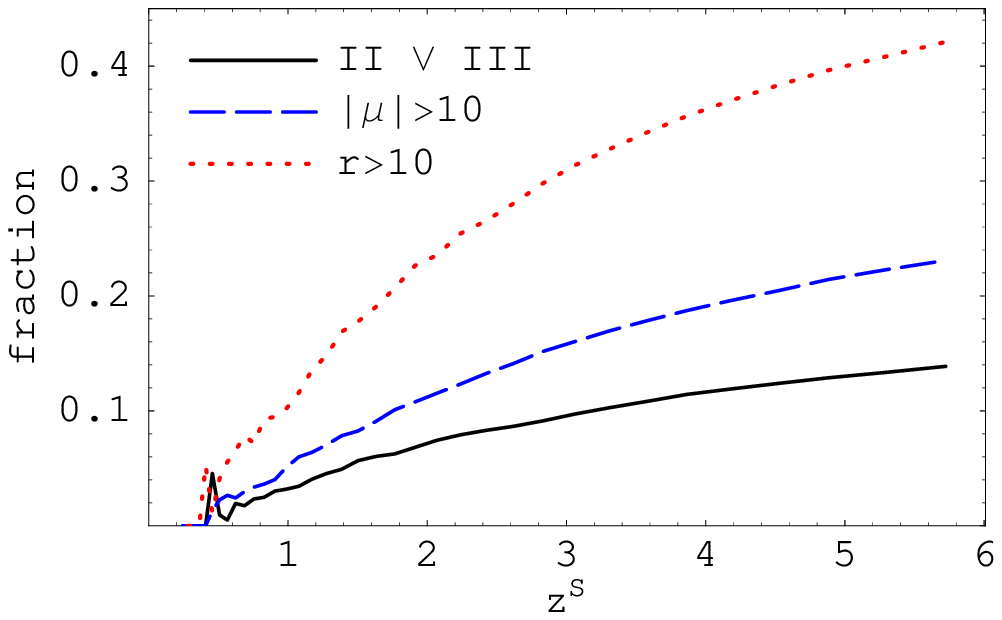}}
\caption{
\label{fig:fraction_no_single_plane}
The fraction of rays of type \IIorIII\ (solid line), of rays with
\mbox{$|\mu|>10$} (dashed line), and of rays with \mbox{$r>10$} (dotted
line) to source redshift $\zS$ for which no single lens plane is able to
generate the relevant property on its own. The relative importance of
foreground and background material clearly increases with increasing $\zS$.}
\end{figure}

For some rays with $\det\matr{A}<0$, $\tr\matr{A}<0$, $|\mu|>10$, or $r>10$,
there is no individual lens plane that is sufficient to produce the relevant
property in the single-plane approximation.  The fraction of rays for which
this is the case is shown in Fig.~\ref{fig:fraction_no_single_plane}. It
increases with increasing source redshift, is lowest for rays of type
\IIorIII, and is highest for rays with $r>10$. This fraction gives an
indication of the extent to which foreground and background material affects
the strong lensing optical depths we have estimated.  Such material
is expected to have no or little effect  on average for the image(s) of a
randomly chosen object of given redshift. However, by selecting rays in the extreme
tail of lensing distributions, we may be significantly biased towards
lines-of-sight for which the additional material enhances the effect of the
primary lens. Fig.~\ref{fig:fraction_no_single_plane} suggests that additional
material is particularly effective in enhancing the probability of highly
distorted images (e.g. giant arcs), presumably because these are sensitive to
the lensing map in a narrow region around its critical lines. 
One should, however, bear in mind that not only do certain directions gain a considered property through the primary lens being supplemented by additional LOS material, but other regions lose the same property because the primary
lens is counteracted by lower than average additional material. Therefore, the fractions of Fig.~\ref{fig:fraction_no_single_plane} do not reflect the overall contribution of line-of-sight material to our cross sections.

\begin{figure}
\centerline{\includegraphics[width=1\linewidth]{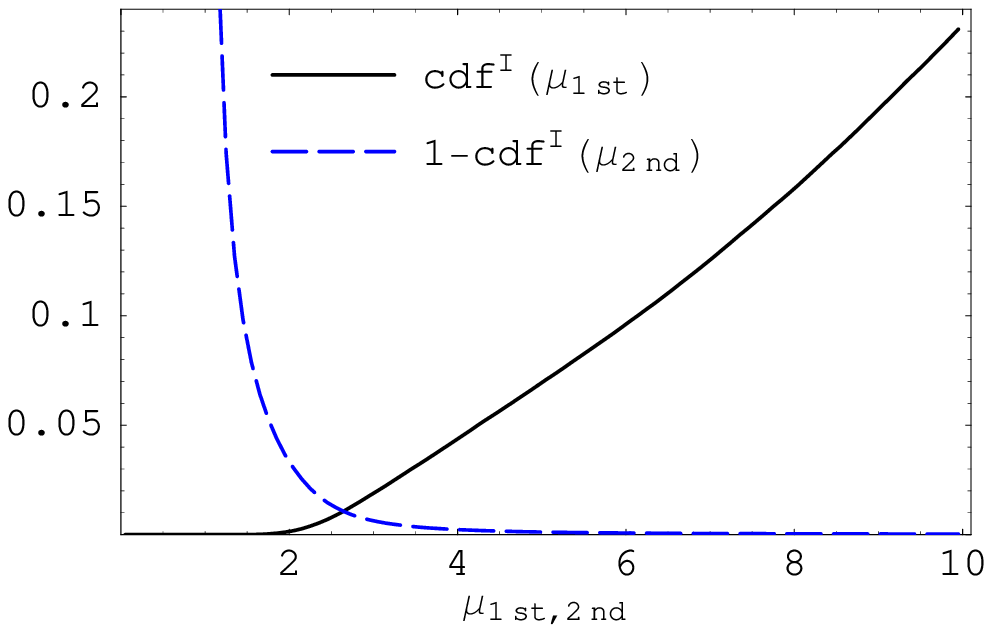}}
\caption{
\label{fig:cdf_of_mu_1st_and_2nd}
The cumulative distribution function $\cdf^\mrm{I}$ of the largest $\mu_\text{1st}$ and
of the second largest $\mu_\text{2nd}$ single-plane magnification for all rays
with a magnification \mbox{$|\mu|>10$} to source redshift $\zS=5.7$. For clarity
we plot $1-\cdf^\mrm{I}(\mu_\text{2nd})$ rather than $\cdf^\mrm{I}(\mu_\text{2nd})$ so that the high tail
of the distribution can be compared with the low tail of $\cdf^\mrm{I}(\mu_\text{1st})$. In
almost all cases the effect due to the primary lens is strongly dominant. }
\end{figure}

In general, the effects of foreground and background material are
relatively weak. In the cases where there is no single plane which is
sufficient to generate the relevant property, there is still usually a single
plane dominating the lensing effects. As an example, we determined for each
ray with $|\mu|>10$ and $\zS=5.7$ the planes which gave rise to the largest
and second largest magnifications, $\mu_\text{1st}$ and $\mu_\text{2nd}$, resp., in the single-plane approximation. The
cumulative distribution of these magnifications is shown
in Fig.~\ref{fig:cdf_of_mu_1st_and_2nd}. Even though the fraction of rays with
$\mu_\text{1st}<10$ is about 23 percent, virtually all rays have
$\mu_\text{1st}>2$ and 93 percent of the rays have $\mu_\text{1st}>5$.  In
most of the cases where there is no sufficient lens plane to cause $|\mu|>10$
alone, there is still an `almost sufficient' plane that gives rise to a
magnification significantly larger than unity. In only 3 percent of cases is
$\mu_\text{2nd}>2$, and in 90 percent of all rays we find $\mu_\text{2nd}<1.4$. Thus
to a good approximation strong lensing can be thought of as being caused by
individual objects. These results agree qualitatively with the findings of
\cite{WaBoOs05} for the distribution of the surface mass density.

\begin{figure}
\centerline{\includegraphics[width=1\linewidth]{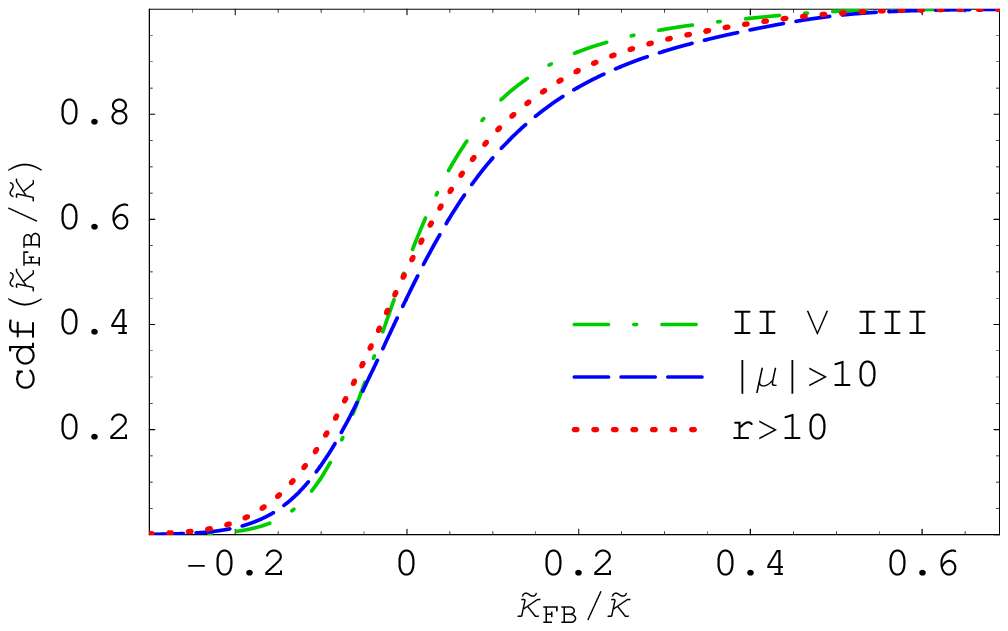}}
\caption{
\label{fig:cdf_rel_kappa_los}
The cumulative distribution function $\cdf(\tilde{\kappa}_\mrm{FB}/\tilde{\kappa})$ of the relative contribution 
$\tilde{\kappa}_\mrm{FB}/\tilde{\kappa}$ of additional matter to the total LOS convergence
$\tilde{\kappa}$ for strongly lensed rays and for source redshift $\zS=5.7$. The dash-dotted line is for rays
of type \IIorIII, the dashed line for rays with \mbox{$|\mu|>10$}, and the
dotted line for rays with \mbox{$r>10$}.}
\end{figure}

There are, nevertheless, a few strongly lensed rays whose properties are due
to more than one object or lens plane. As noted above, two or more objects at
similar redshift contribute significantly for a few percent of all rays.
Fig.~\ref{fig:cdf_of_mu_1st_and_2nd} shows that for a further few percent two or more
uncorrelated objects at different redshifts make a significant
contribution. For the remaining rays the overall effects of foreground and
background matter are at a much lower level. To demonstrate this
quantitatively, we first determined the projected mass overdensity at the
position of each ray on each plane. We then divided these overdensities by the
critical surface densities of the relevant planes. Finally, for each ray we
summed the contributions from all planes to obtain a `LOS convergence'
$\tilde{\kappa}$, which -- to be more precise -- is the
lensing-efficiency-weighted projection of the matter overdensity along the
ray. (This is equal to the convergence $\kappa$ in the single-plane and
weak-lensing approximations.)  We then performed a similar calculation for all
rays of type \IIorIII, with $|\mu|>10$ or with $r>10$. In addition, when calculating $\tilde{\kappa}$ for these rays, we excluded the contribution of the plane containing the primary lens to isolate the contribution $\tilde{\kappa}_\mrm{FB}$ of foreground and background matter to the lensing event.
The cumulative distribution of the ratio of $\tilde{\kappa}_\mrm{FB}$ to  $\tilde{\kappa}$ is shown in Fig.~\ref{fig:cdf_rel_kappa_los}. For 20 percent of the rays of type \IIorIII, additional matter along the LOS contributes more than 10 percent to the total LOS convergence. On the other hand, for 50 percent of the rays, there is a negative contribution $\tilde{\kappa}_\mrm{FB}$. Although there is a noticeable fraction of strongly lensed rays whose LOS convergence is enhanced by additional matter along the LOS, there is also a noticeable fraction of strongly lensed rays whose LOS convergence is decreased due to the lack of matter along the LOS.

\begin{figure}
\centerline{\includegraphics[width=1\linewidth]{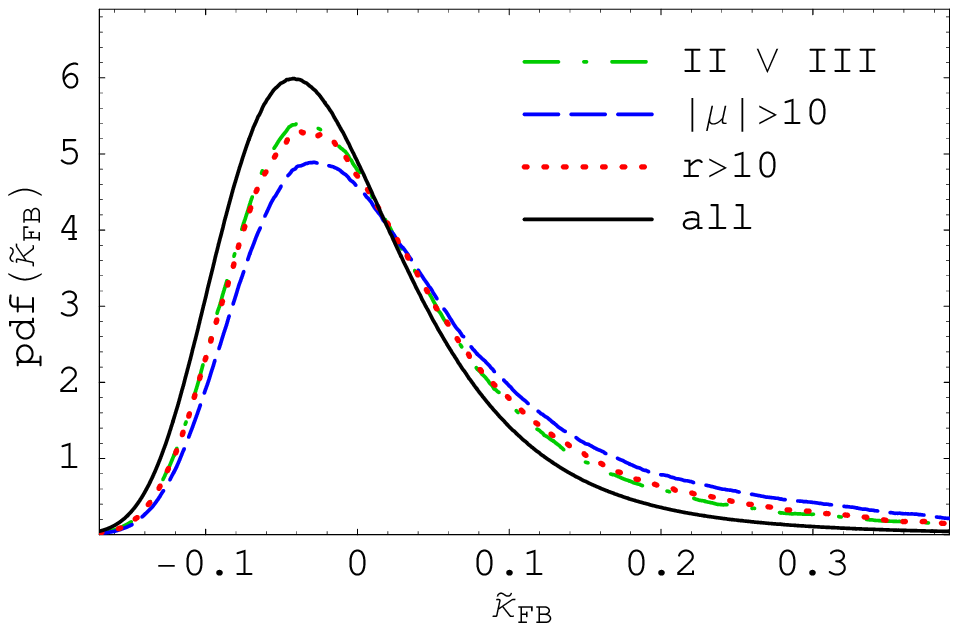}}
\caption{
\label{fig:pdf_kappa_los}
The probability density function $\pdf(\tilde{\kappa}_\mrm{FB})$ of the LOS convergence
$\tilde{\kappa}_\mrm{BF}$ for strongly lensed rays and for source redshift $\zS=5.7$,
but with the primary lens contribution excluded. The dash-dotted line is for rays
of type \IIorIII, the dashed line for rays with \mbox{$|\mu|>10$}, and the
dotted line for rays with \mbox{$r>10$}. For comparison, the solid line
shows the corresponding distribution $\pdf(\tilde{\kappa})$ for all rays irrespective of their
lensing properties. A small but significant shift towards larger convergence
is visible in the direction of strong-lensing events. }
\end{figure}

\begin{figure}
\centerline{\includegraphics[width=1\linewidth]{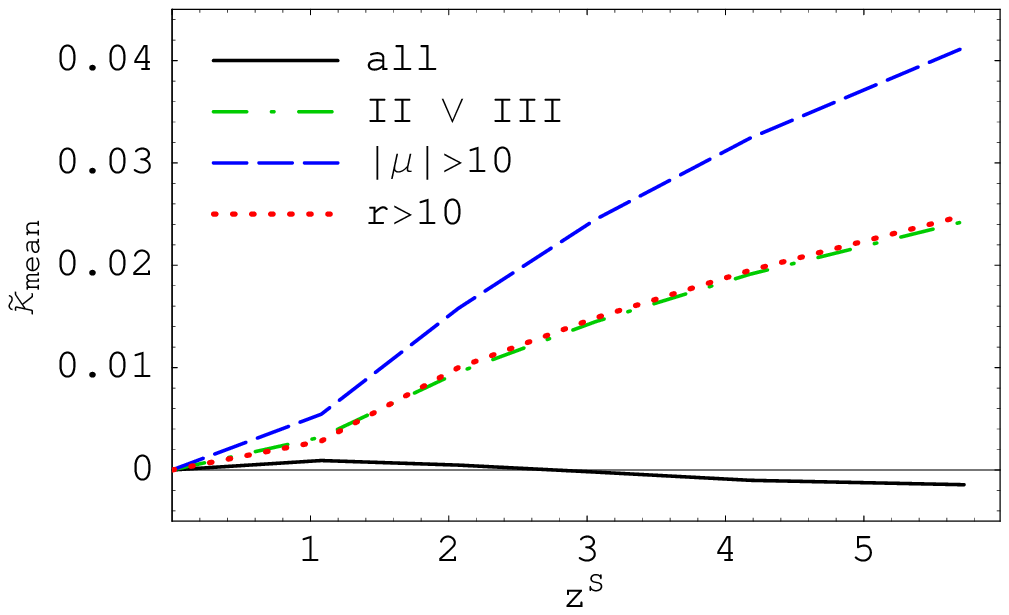}}
\caption{
\label{fig:mean_kappa_los_of_zS}
The mean $\tilde{\kappa}_\mrm{mean}$ of the distributions shown in
Fig.~\ref{fig:pdf_kappa_los} as a function of source redshift $\zS$. The line
types correspond to those of Fig.~\ref{fig:pdf_kappa_los}. The mean
LOS convergence for all rays irrespective of their lensing properties is 
non-zero only because of residual sampling variance. The mean convergence
from foreground and background matter along rays containing a strong lens
is significantly non-zero, however, although still much smaller than the
typical convergence due to the primary lens. }
\end{figure}

In Fig.~\ref{fig:pdf_kappa_los}, distributions of the LOS convergence for
$\zS=5.7$ are compared for all rays and for strongly lensed rays with the
primary lens contribution removed.  Although the distributions are very
similar, small shifts are visible.
In Fig.~\ref{fig:mean_kappa_los_of_zS}, we
show the means of these distributions as a function of source redshift
$\zS$. By definition, the mean LOS convergence of all LOS should be zero. The
measured mean for our whole ray sample is not exactly zero because of sampling
variance,\footnote{
One might naively expect the sampling variance to be negligible for the 640 million rays we shot. However, the rays were not shot independently, but are confined to forty different patches of $40\times40\,\Mpc^2/h^2$ on each lens plane. The matter content of each patch is still subject to significant cosmic variance, and so, therefore, is the combined sampling area.
} but it is much smaller than the mean for strongly lensed rays with
the primary lens excluded. This demonstrates a small but measurable bias
towards selecting directions in which matter in front or behind the primary lens
enhances the lensing. The effect increases with increasing source redshift, is
weakest for the sample of rays with $r>10$, and is strongest for $|\mu|>10$.
For $\zS=5.7$ and $|\mu|>10$, where the bias is strongest, we find an average contribution of 0.04 to the LOS convergence.
Hence, in all cases the bias is small in comparison with the effect of the primary lens for
which $\tilde{\kappa}\sim 1$.

\section{Summary}
\label{sec:summary}
The aim of this work has been to study the statistical distribution of the
distortion of images of distant sources due to gravitational lensing. In
particular, we have concentrated on estimating the cross section for rare
strong-lensing events.  Our results were obtained by shooting random rays
through a series of lens planes created from the Millennium
Simulation~\citep{SpringelEtAl05MSReview}. This is the largest $N$-body
simulation of cosmological structure formation available today. We have
devised improved algorithms to make the lens-planes and to calculate bending
angles and shear matrices on these planes in order to take full advantage of
the very large volume and the high spatial and mass resolution offered by the
simulation.

In Sec.~\ref{sec:magnification_distribution}, we presented results for the
statistical distribution of the magnification of point sources.  The
distribution is skewed with a peak at magnifications below unity and a tail
toward high magnification. With increasing source redshift, the peak broadens
and moves to lower magnifications, while the tail gains more weight.  The
magnification distribution affects the observed luminosity distribution of
astronomical standard candles. For type Ia supernovae, perhaps the most
interesting case, magnification effects on the luminosity distribution are
still small for current samples compared to the intrinsic luminosity scatter,
to extinction and to other effects. In future high-redshift, high-precision
surveys, however, such magnification effects may cause significant systematic
errors, so it will be necessary to detect and to correct for them
\citep{DoVa06,MuVa06astroph}. In the most optimistic case, detailed comparison
with predictions of the magnification distribution may help to discriminate
between cosmological models.

Various optical depths connected to strong lensing were presented in
Sec.~\ref{sec:optical_depths}. In particular, we estimated the fraction of
images of sufficiently small sources that are highly magnified, have a large
length-to-width ratio, or belong to multiply imaged sources. All the optical
depths we analyse increase strongly with increasing redshift. In comparison
with earlier results by, e.g., \cite{LiEtal05} and \cite{WaBoOs04}, we find a
stronger evolution with source redshift, leading to higher optical depths for
source redshifts $\zS>1$. We discussed possible reasons for this difference.

The results we presented in Sec.~\ref{sec:lens_properties} show that
significant contributions to the strong-lensing optical depths come from dark
matter halos with masses between $10^{13}h^{-1}\,\Msolar$ and $10^{15}h^{-1}\,\Msolar$.
The upper mass limit is due to the very rapidly decreasing abundance of more
massive structures. This exponential decrease occurs at lower mass at higher
redshift, and in conjunction with the lens geometry it explains why almost all
lenses are at redshifts $\zL<2.5$, even for sources with $\zS>5$.  The lower
mass limit for strong lensing is due primarily to the small cross sections of
individual low-mass halos, although the spatial and mass resolution limits of
the simulation itself and of our lens planes also play some role.  Estimates
based on analytic results for spherical NFW halos fit to the Millennium data
suggest that these resolution effects are subdominant, and probably only
reduce our total cross sections by of order 15\%. A more important effect on
the relevant scales is our neglect of the baryonic mass of the central
galaxies.  We will come back to this in later work.

We find that the mass range over which halos can cause strong lensing extends
to lower masses than those given by \cite{LiEtal05} and \cite{DaHoHe04}. This
difference may in part reflect the lower resolution of the simulations used by
these authors, and in part the fact that they considered extended sources with
diameters $\sim 0.1$--$1\,\arcsecs$, while we assumed sufficiently small sources when
calculating our cross sections.  Halos near our lower mass limit have Einstein
radii of order $1\,\arcsecs$ and so are inefficient in producing strongly distorted
images of sources of comparable angular extent.

Since our set of lens planes represents the whole matter distribution between
source and observer, we are able to quantify the influence of foreground and
background matter on the frequency and the properties of strong lensing
events. We find that such effects are quite modest. On average, the contribution of foreground and background material is
only a few percent. Although we do find a bias towards excess
foreground and background matter on strong-lensing lines-of-sight, the effect
is significantly smaller than suggested by \cite{WaBoOs05}.

The most obvious extension to the work we have presented here would be
ray-tracing studies of the effects of lensing for realistic
distributions of source properties and across finite size fields. This will,
for example, allow direct comparison with observations of massive galaxy
clusters where many sets of multiple images are now detected in the best
cases \citep{BroadhurstEtal05,HaSePa06}. When galaxy properties from galaxy formation modelling within the
Millennium Simulation \citep[e.g.][]{SpringelEtAl05MSReview,CrotonEtal06,BowerEtal06,LuBl07} are combined with such ray-tracing
analyses, it will be possible to study whether the dark halo masses of
individual cluster galaxies are consistent with observation, providing an
additional observational test of the hierarchical build-up of structure
predicted by the standard $\Lambda$CDM model. This combination of
semi-analytic simulation of galaxy formation with ray-tracing measures of 
lensing will also allow an estimate of how our strong-lensing cross sections
should be modified to account for the galaxies, as well as detailed studies
of predictions for galaxy-galaxy lensing. We intend to come back to all these
topics in future work.

\section*{Acknowledgments}
 We thank Volker Springel and Jeremy Blaizot for helpful
discussions concerning the software development.
Furthermore, we thank Ole M{\"o}ller, Matthias Bartelmann, Joachim Wambsganss, Guoliang Li, and Shude Mao for helpful discussions concerning
gravitational lensing. This work was supported by the DFG within the Priority Programme 1177
under the projects SCHN 342/6 and WH 6/3.

\appendix

\section{Integral relations for the magnification}
\label{sec:integrals_of_mu}

For a well behaved non-singular lens system, the numbers of images of type I,
II and III of a given source satisfy $n_\mrm{I} - n_\mrm{II} + n_\mrm{III} =
1$ \citep{ScEhFa92}.  Here we briefly discuss an `integral version' of this
theorem for the particular geometry used in our work: the Multiple-Plane
Approximation with lens planes carrying a smooth and non-singular matter
distribution that is periodic with a rectangular unit cell of dimensions $L_1
\times L_2$. In this case, the image plane and source plane can both be
represented by a rectangle $\mbb{P}=[0,L_1]\times[0,L_2]$ with periodic
boundary conditions. Furthermore, the lens mapping
\begin{equation*}
 \mcal{L}: \mbb{P} \rightarrow \mbb{P}: \vect{\theta}\mapsto\vect{\beta}(\vect{\theta})=\vect{\theta}+\vect{\alpha}(\vect{\theta})
\end{equation*}
from image position $\vect{\theta}$ to source position $\vect{\beta}$ for a given source redshift is then smooth and
non-singular with a smooth, non-singular, and periodic deflection angle $\vect{\alpha}(\vect{\theta})$. Under these conditions, the inverse signed magnification
$\mu^{-1}=\det(\partial\vect{\beta}/\partial\vect{\theta})$ satisfies the
following relation:
\begin{equation}
\label{eq:theorem_inverse_mu_integral}
1= \frac{1}{||\mbb{P}||}\int_{\mbb{P}} \mrm{d}^2\vect{\theta}\; \mu^{-1}(\vect{\theta}),
\end{equation}
where $||\mbb{P}||=L_1 L_2$ denotes the area of the rectangle $\mbb{P}$. The
following derivation employs integration by parts and exploits the smoothness and periodicity of
$(\partial\vect{\beta}/\partial\vect{\theta})$:
\begin{equation*}
 \begin{split}
\int_{\mbb{P}} \mrm{d}^2\vect{\theta}\; \mu^{-1}(\vect{\theta})
 &=
\int_{\mbb{P}} \mrm{d}^2\vect{\theta}\;\det\left(\parder{\vect{\beta}}{\vect{\theta}}\right)
\\&=
\int_0^{L_1} \mrm{d}\theta_1 \int_0^{L_2} \mrm{d}\theta_2\;
\left(\parder{\beta_1}{\theta_1}\parder{\beta_2}{\theta_2} - \parder{\beta_1}{\theta_2}\parder{\beta_2}{\theta_1}\right)
\\&=
\int_0^{L_2} \mrm{d}\theta_2\left( \int_0^{L_1} \mrm{d}\theta_1\;
\parder{\beta_1}{\theta_1}\parder{\beta_2}{\theta_2} \right)
\\&-
\int_0^{L_1} \mrm{d}\theta_1 \left( \int_0^{L_2} \mrm{d}\theta_2\;
\parder{\beta_1}{\theta_2}\parder{\beta_2}{\theta_1} \right)
\\&=
\int_0^{L_2} \mrm{d}\theta_2\left(
\left[\beta_1\parder{\beta_2}{\theta_2}\right]_{\theta_1=0}^{L_1}
-\int_0^{L_1} \!\mrm{d}\theta_1\;\beta_1\parder{^2\beta_2}{\theta_1\partial\theta_2}
\right)
\\&-
\int_0^{L_1} \mrm{d}\theta_1 \left(
\left[\beta_1 \parder{\beta_2}{\theta_1}\right]_{\theta_2=0}^{L_2}
-\int_0^{L_2} \!\mrm{d}\theta_2\;
\beta_1\parder{^2\beta_2}{\theta_1\partial\theta_2}
\right)
\\&=
\int_0^{L_2} \mrm{d}\theta_2
\left[\beta_1\parder{\beta_2}{\theta_2}\right]_{\theta_1=0}^{L_1}
-
\int_0^{L_1} \mrm{d}\theta_1 
\left[\beta_1 \parder{\beta_2}{\theta_1}\right]_{\theta_2=0}^{L_2}
\\&=
\int_0^{L_2} \mrm{d}\theta_2\left[ \beta_1(L_1,\theta_2)-\beta_1(0,\theta_2)\right]\parder{\beta_2(L_1,\theta_2)}{\theta_2}
\\&-\int_0^{L_1} \mrm{d}\theta_1\left[ \beta_1(\theta_1,L_2)-\beta_1(\theta_1,0)\right]\parder{\beta_2(\theta_1,L_2)}{\theta_2}
\\&=
\int_0^{L_2} \mrm{d}\theta_2 L_1\parder{\beta_2(L_1,\theta_2)}{\theta_2}
\\&=
L_1 L_2\;.
 \end{split}
\end{equation*}

For our ray sampling method, it follows directly from
relation~\eqref{eq:theorem_inverse_mu_integral} that a representative sample
of rays with random positions $\vect{\theta}_i$ ($i=1,\ldots,N$) in the image
plane should satisfy:
\begin{equation}
\label{eq:mu_sample_sum}
\frac{1}{N}\sum_{i=1}^N\mu^{-1}(\vect{\theta}_i)\approx  \frac{1}{||\mbb{P}||}\int_{\mbb{P}} \mrm{d}^2\vect{\theta}\; \mu^{-1}(\vect{\theta})=1.
\end{equation}
For our ray sample, we find that
\begin{equation*}
\left| \frac{1}{N}\sum_{i=1}^N\mu^{-1}(\vect{\theta}_i) \right| - 1 < 0.003
\end{equation*}
for all source redshifts.

For the magnification distribution, Eq.~\eqref{eq:theorem_inverse_mu_integral}
implies that
\begin{equation*}
\begin{split}
\mathrm{pdf}^\mrm{S}(\mu')
&=
\totder{}{\mu'}\frac{\int_{\mbb{P}}\diff[2]{\vect{\theta}}\left|\mu(\vect{\theta})\right|^{-1} \Theta[\mu'-\mu(\vect{\theta})]}
{\int_{\mbb{P}}\diff[2]{\vect{\theta}}\left|\mu(\vect{\theta})\right|^{-1}}
\\&=
 \frac{\int_{\mbb{P}}\diff[2]{\vect{\theta}}\left|\mu(\vect{\theta})\right|^{-1} \delta[\mu'-\mu(\vect{\theta})]}
{\int_{\mbb{P}}\diff[2]{\vect{\theta}}\left|\mu(\vect{\theta})\right|^{-1}}
\\&=|\mu'|^{-1}
\frac{\int_{\mbb{P}}\diff[2]{\vect{\theta}} \delta[\mu'-\mu(\vect{\theta})]}
{\int_{\mbb{P}}\diff[2]{\vect{\theta}}\left|\mu(\vect{\theta})\right|^{-1}}
\\&=\frac{\int_{\mbb{P}}\diff[2]{\vect{\theta}}}{\int_{\mbb{P}}\diff[2]{\vect{\theta}}\left|\mu(\vect{\theta})\right|^{-1}}
|\mu'|^{-1}\mathrm{pdf}^\mrm{I}(\mu')
\\&=\left(1-2\,\tau^\mrm{S}_\mrm{II}\right)|\mu'|^{-1}\mathrm{pdf}^\mrm{I}(\mu')
\;.
\end{split}
\end{equation*}
In practice, the optical depth $\tau^\mrm{S}_\mrm{II}$ for images with negative magnification is very small. Hence,
\begin{equation}
\label{eq:pdf_mu_relations}
\begin{split}
\mrm{pdf}^\mrm{S}(\mu')&\approx|\mu'|^{-1}\mrm{pdf}^\mrm{I}(\mu').
\end{split}
\end{equation}
Employing the fact that both probability distributions are normalised, we 
finally find:
\begin{equation}
\label{eq:pdf_mu_integrals}
\int_\mbb{R}\left|\mu\right|^{-1}\mathrm{pdf}^\mrm{I}(\mu)\,\diff\mu\approx 1
\;,\;\text{ and}\;\;\;
\int_\mbb{R}\left|\mu\right|\mathrm{pdf}^\mrm{S}(\mu)\,\diff\mu\approx 1
.
\end{equation}

\section{Empty-beam magnification in a flat universe}
\label{sec:empty_beam_magnification}

The Jacobian of the lens mapping for light propagation through an 
inhomogeneous universe is given by \citep[see, e.g.,][]{Schneider06_SaasFee33Part1}:
\begin{multline}
\label{eq:CS_Jacobian}
\tens{A}_{ij}(\vect{\theta},w) = \delta_{ij} \\- \frac{2}{c^2}\int_0^w\!\!\diff{w'}
\frac{(w-w')w'}{w}\frac{\partial^2\Phi(\vect{x}(\vect{\theta}, w'),w')}
{\partial \theta_i \,\partial\theta_k}\, \tens{A}_{kj}(\vect{\theta},w').
\end{multline}
Here, $\Phi$ denotes the three-dimensional gravitational potential, $w$ the comoving line-of-sight distance and $\vect{\theta}$ the direction of the light ray, and $\vect{x}$ the comoving transverse separation. For an empty beam, we have
$\tens{A} = {\rm diag}(\mu_\mrm{min}^{-1/2},\mu_\mrm{min}^{-1/2})$. Furthermore, the Poisson equation in this case reads
\begin{equation*}
 \nabla^2 \Phi(\vect{x}(\vect{\theta}, w),w) = -\frac{3H_0^2\Omega_{\rm m}}{2a(w)}\;,
\end{equation*}
where $a(w)$ is the scale factor.

Considering now the $1,1$-component of Eq.~\eqref{eq:CS_Jacobian}, we find
\begin{equation*}
\mu_\mrm{min}^{-1/2}(w) = 1 + \frac{3H_0^2\Omega_{\rm m}}{2c^2}\int_0^w\!\diff{w'}\frac{(w-w')w'}{a(w')w}\,\mu_\mrm{min}^{-1/2}(w')\; .
\end{equation*}
Differentiating this expression twice, we finally obtain the differential 
equation 
\begin{equation*}
\frac{\diff[2]{f}}{\diff{w^2}} = \frac{3H_0^2\Omega_{\rm m}}{2c^2}\,\frac{f(w)}{a(w)}\;,
\end{equation*}
where $f(w)\equiv w\mu_\mrm{min}^{-1/2}(w)$. This ordinary differential equation can be easily solved numerically.

\bibliographystyle{aa}

\end{document}